\documentclass[10pt,conference,compsocconf]{IEEEtran}

\pdfoutput=1
\newif\ifcameraready
\ifcameraready \else\IEEEoverridecommandlockouts \fi
\ifcameraready\else\fi
\usepackage[utf8]{inputenc}
\usepackage{amsmath,amssymb,mathtools}

\usepackage{enumitem}
\usepackage{framed}
\usepackage{fixltx2e}
\usepackage[numbers,sort,compress]{natbib}
\newtheorem{proposition}{Proposition}
\newtheorem{lemma}{Lemma}
\newtheorem{corollary}{Corollary}
\newcommand{\sgn}{\operatorname{sgn}}
\newcommand{\ang}[1] {\langle #1 \rangle}

\newcommand{\hide}[1]{} 
\newcommand{\Section}{Sec.}
\newcommand{\Sections}{Sec.}
\newcommand{\E}{\mathcal{E}}
\newcommand{\Ps}{\mathcal{P}}
\newcommand{\A}{\mathcal{A}}
\newcommand{\BR}{\mathrm{BR}}
\newcommand{\Li}{\mathcal{L}}
\newcommand{\pos}{\mathrm{pos}}
\newcommand{\supp}{\mathrm{supp}}
\newcommand{\Perm}{\mathrm{Perm}}
\newcommand{\unif}{\mathrm{unif}}
\renewcommand{\vec}[1]{\boldsymbol{#1}}

\begin{document}
\ifcameraready
\else\title{Picking vs. Guessing Secrets: A Game-Theoretic Analysis\\(Technical Report)\thanks{This manuscript is the extended version of our conference paper \cite{khouzani15picking}.}}\fi
\author{\IEEEauthorblockN{MHR Khouzani\IEEEauthorrefmark{1},
Piotr Mardziel\IEEEauthorrefmark{2},
Carlos Cid\IEEEauthorrefmark{3},
Mudhakar Srivatsa\IEEEauthorrefmark{4}}
\IEEEauthorblockA{\IEEEauthorrefmark{1}Queen Mary, University of London}
\IEEEauthorblockA{\IEEEauthorrefmark{2}University of Maryland, College Park}
\IEEEauthorblockA{\IEEEauthorrefmark{3}Royal Holloway, University of London}
\IEEEauthorblockA{\IEEEauthorrefmark{4}IBM T.J. Watson Research Laboratory}}
\maketitle
\begin{abstract}
 Choosing a hard-to-guess secret is a prerequisite in many security 
applications. Whether it is a password for user authentication or a secret key 
for a cryptographic primitive, picking it requires the user to trade-off 
usability costs with resistance against an adversary: a simple password is 
easier to remember but is also easier to guess; likewise, a shorter 
cryptographic key may require fewer computational and storage resources but it 
is also easier to attack.  A fundamental question is how one  
can optimally resolve this trade-off. 
A big challenge is the fact that an adversary can also utilize the knowledge 
of such usability vs. security trade-offs to strengthen its attack. 

In this paper, we propose a game-theoretic framework for analyzing the optimal 
trade-offs in the face of strategic adversaries. We consider two types of 
adversaries: those limited in their number of tries, and those that are ruled by 
the cost of making individual guesses. For each type, we derive the 
mutually-optimal decisions as Nash Equilibria, the strategically pessimistic 
decisions as maximin, and optimal commitments as 
Strong Stackelberg Equilibria of the game.
We establish that when the adversaries are faced with a capped number of 
guesses,  the user's optimal trade-off  is 
a uniform randomization over a subset of the secret domain.
On the other hand, when the attacker strategy is ruled by the cost of making 
individual guesses, Nash Equilibria may completely fail to provide 
the user with any level of security, signifying the crucial role of credible 
commitment for such cases. We illustrate our results using numerical examples 
based on real-world samples and discuss some policy implications of our work.

\end{abstract}
\section{Introduction}
Passwords remain the most common means of authenticating humans to 
computer systems. Yet, passwords are also among the most common 
points of failure of security systems 
\cite{bonneau2012quest,bonneau2012guessing,florencio2014administrator}. 
According to an investigation report in 2011 \cite{team2008data}, stolen login 
credentials accounted for nearly a third of 
corporate data breach incidents, out of which, more than a quarter were 
estimated to be carried out using a form of a guessing attack. 
Poorly chosen passwords undermine an otherwise secure authentication system.
Users tend to choose easy to remember passwords
\cite{sasse2001transforming,yan2004password}.  This is rationalizable as 
attempts in balancing usability costs with perceived security. 

Managing the utility vs. security trade-off is also a relevant problem
in the application of cryptographic techniques, which usually rely on
maintaining a key unknown to any adversary.
Longer keys provide stronger security guarantees but at the same time
inflict larger storage and computational costs on the system.
Using cryptographic techniques, therefore, entails trading off
utility for security, either through the choice of key size or the
method of key generation.
Either way, the decision must be made in the context of adversaries.

Guessing attacks are often categorized as \emph{online} and \emph{offline} based 
on their context of execution \cite{boyen2007halting}.
Online attacks involve interacting with the target system.
In such an attack, adversaries are often limited in the number of
(failed) guesses they can make (within a certain time period) before the system 
prevents any further interaction.
In the case of password authentication, this is usually an account
lock-out that requires intervention of the legitimate user using an
alternate channel of authentication (email, phone, etc.).

In \emph{offline} attacks, adversaries are assumed to have collected 
sufficient data to examine unlimited number of guesses, and are only
constrained by their computational resources. 
In the case of password authentication, for example, this data can be 
the leaked hashes of user passwords, enabling the attackers to compute hashes 
of their guesses and compare them for a match, theoretically an unlimited 
number of times. 
Another example of an offline attack setting is when an adversary eavesdrops a 
cryptographic response to a predictable challenge in a 
challenge-response authentication protocol.
Although unlimited in the number of guesses, adversaries in such offline 
scenarios still need to be wary of costs of trying guesses as 
computation of password hashes or cryptographic responses are not 
instantaneous or free (specially, noting that hash functions for 
hashing passwords are intentionally chosen to be slow on hardware to dissuade 
brute-force attacks). Hence, the response of such adversaries is governed by 
the computational/time cost per each guess. 
An adversary may obtain a pre-computed list of hashes to remove (or a 
\emph{rainbow} table to mitigate) the computational burden during the execution  
of the attack. In such cases, 
the bottle-neck becomes the storage requirement for such a table, which implies 
a cap on the number of available guesses, similar to the online case.

We will collectively refer to passwords or cryptographic keys as 
\emph{secrets}. 
We also use the terms \emph{Capped-Guesses} and \emph{Costly-Guesses} to 
respectively describe  the following two settings: 
(1)~adversaries are limited in their number of guesses, e.g. in online password 
attacks in the presence of a rate limiting mechanism, or in
offline attacks that use storage-limited pre-computed tables; and 
(2)~adversaries incur a cost per each guess, e.g. in brute-force offline 
attacks.
Regardless of the type of the guessing attack, the inherent behavioral or 
systematic preferences over the secret space can be exploited by adversaries and 
boost their guessing efficiency.
Therefore, any secret picking policy that aims to achieve a desirable trade-off 
between usability and security must evaluate the possible 
reaction of a rational adversary given their capabilities.
In particular, it is insufficient to analyze the decisions of either the users 
or the adversaries without taking into account the reaction of the other.
Game theory provides tools to analyze such 
strategic interactions. The notion of \emph{equilibrium}, in particular, 
describes how rational parties would eventually behave when faced
against each other by characterizing their mutually-optimal strategies. 

The basic question at the heart of this paper is the following: 
given a known uneven usability cost over the space of secrets, how can the 
defender optimally randomize in picking a secret? 
The main contribution of the paper is answering this fundamental question.
Specially:
\begin{itemize}[noitemsep,nolistsep,leftmargin=*]
\item{} We present novel decision and game-theoretic models for both 
Capped-Guesses and Costly-Guesses settings that are simple enough to allow 
analysis yet general enough to cover all the cases described above.
\item{} We provide complete analysis of these games and discuss the security 
implications of the solutions. Specifically, we derive optimal secret selection 
policies with respect to different strategic metrics, namely, the strategically 
pessimistic solutions (Maximin), the mutual-best-response solutions (Nash 
Equilibria -- NE), and the optimal commitment strategies (Strong Stackelberg 
Equilibria -- SSE).
\item{} For Capped-Guesses settings, we show that, interestingly, the 
optimal picking strategies still constitute uniform distributions despite the 
uneven preferences of the picker over the secret space. 
The trade-off is achieved by randomizing only over a (lower cost) subset of the 
secret space, while the probability distribution over the subset is 
uniform. The size of the subset is influenced by the picker's trade-off parameters and (only) the cap on the available guesses. 
The optimal guessing strategies are restricted to the same subset though they are not uniform. 
Instead, the guesser probes the picker's more favored secrets in that subset with higher probabilities. 
We also show that for this scenario, all of the different strategic metrics of 
Maximin, NE and SSE lead to the same solution for the picker.
\item{} For the Costly-Guesses settings, we find a surprising result, 
reminiscent of the prisoner's dilemma situation: aside from 
trivial cases, the NE strategies of the picker fail to yield any desirable 
security level, irrespective of the size of the secret space or the cost 
associated with the loss of the secret. We demonstrate how the picker can 
retrieve a desirable usability-security trade-off using \emph{commitment} to 
optimal randomizations.  We also notice that these optimal commitment (SSE) 
strategies for this case are almost never completely uniform, though they resemble uniform selection,
with diminishing tails on costlier secrets. 
\item{} We provide numerical illustrations of our analyses using examples such 
as the leaked RockYou password dataset and cryptographic keys with increasing 
costs in their size.
\end{itemize}

The paper is structured as follows: \Section~\ref{sec:Model} introduces the 
building blocks of our \emph{non-zero-sum} \emph{two-player} game between a 
picker and a guesser.
In \Section~\ref{sec:model:capped_guesses}, we present the model for 
Capped-Guesses scenarios and introduce different game theoretic notions of a 
solution, which we fully derive in \Section~\ref{sec:Analysis:capped_guesses}.  
In \Sections \ref{sec:model:costly_guesses} and 
\ref{sec:Analysis:costly_guesses}, we present the model and analysis of the 
Costly-Guesses scenarios. 
In \Section~\ref{sec:Discussions} we comment on some of the implications of our 
results. A brief overview of related literature is discussed in 
\Section~\ref{sec:Related_Work}. A summary of our results and some suggestions 
for future directions of research concludes our paper in 
\Section~\ref{sec:Conclusion}. 
\ifcameraready
\else
The technical proofs of the results are all aggregated in the Appendices of this technical 
report.\fi


\section{Model}\label{sec:Model}
In what follows, we progressively construct the model of our non-zero-sum 
two-person games between the \emph{picker}, and the \emph{guesser}. Critically, 
we assume that the parameters of the games are ``common knowledge'', i.e., both 
players are aware of the presence and type of the game, the utilities and the 
information available to each other. 

The picker (she) chooses a \emph{secret} from the finite set of all secrets 
$\Ps=\{p_1,\ldots,p_{|\Ps|}\}$. 
Let $d\in\Ps$ denote a  pure (i.e., deterministic) \emph{action} of the picker.
$\Ps$ is thus the picker's \emph{pure action set}. 
The picker has uneven preferences over this set of secrets. In the case of password selection, for instance, 
this preference  could be related to the memorability and 
ease of use: simpler passwords are easier to remember and less cumbersome to 
type in.  In the spirit of the von Neumann-Morgenstern utility theorem 
\cite{kuhn2007introduction}, we model these preferences by assigning different 
costs to different secrets.\footnote{Note, however, that we assume the 
usability costs and security costs of the picker are additive through an 
appropriate scaling.} Specifically, let the whole set of secrets be 
partitioned into disjoint non-empty subsets $\mathcal{E}_1$,
\ldots, $\mathcal{E}_N$, i.e., $\E_i\neq\emptyset$ for all $i$, $\E_i\cap 
\E_j=\emptyset$ for $i\neq j$ and $\cup_{i=1}^N \mathcal{E}_i=\mathcal{P}$, such 
that the picker incurs a usability cost of $C_i$ if she picks any of the 
members of the set $\mathcal{E}_i$ as her secret. Without loss of generality,
assume $0\leq C_1<\ldots<C_N$. Hence, in the absence of any 
other considerations, the picker prefers to choose her secret from set 
$\mathcal{E}_i$ rather than $\mathcal{E}_j$ when $i<j$, as she assigns a lower 
usability cost to secrets from the first set. These data are determined, for 
instance in the case of password choice, by statistical investigation of the 
past databases of cracked passwords, e.g., as published in 
\cite{bonneau2012science,zviran1999password}. 
Alternatively, these sets can represent passwords that minimally 
satisfy an \emph{increasingly more complex} password creation rule-sets. For 
instance, $\mathcal{E}_1$ can be the set of all dictionary words in lower case, 
$\E_2$ the set of all dictionary words but requiring a mix of capital 
letters, $\E_3$ having the additional rule of including a number as well, 
$\E_4$ requiring a symbolic character too,  {etc}.\footnote{It is natural to 
assume that these partitions are common knowledge as \emph{sets} as opposed to 
\emph{lists}. In particular, no specific indexing of the members of a partition 
is common knowledge, and hence, the solutions must be \emph{symmetric within the 
partitions}. Nevertheless, we provide our analysis agnostic of any assumption 
about existence of a common indexing inside partitions. Symmetric solutions can 
then be extracted.}

The guesser (he) makes guesses about the choice of the picker. Upon the 
discovery of the secret, i.e., a correct guess, the guesser wins a gain of 
$\gamma>0$, and the picker incurs a loss of $\lambda>0$. 
The guesses are either constrained in number or subject to cost. We will 
investigate these two cases separately in 
Sections~\ref{sec:model:capped_guesses} and \ref{sec:model:costly_guesses} 
respectively. 
In what follows, we provide 
two numerical instances of the model. Note that these numerical examples 
are mainly for the purpose of illustrating the analysis.

\subsection{Cost Example:  Passwords}\label{sec:example-password-model}
\begin{figure}%
\includegraphics[width=0.95\linewidth]{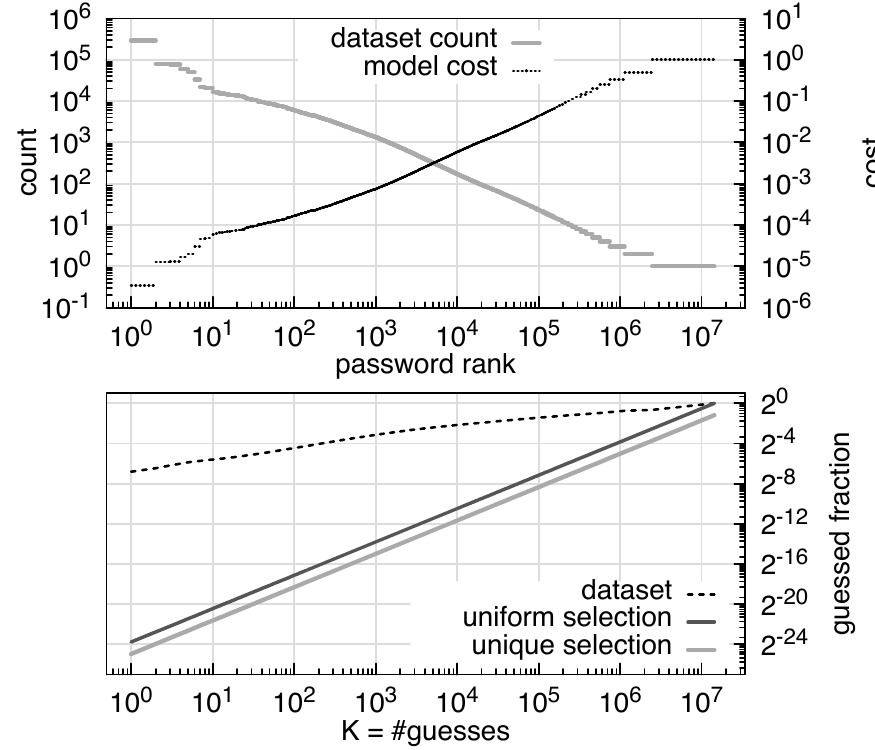}
\caption{\label{fig:pass-dataset}(Top) RockYou dataset
  password frequency and the derived cost model.
  (Bottom) Fraction of passwords guessed as a function of
  number of guesses.}
\end{figure}%

The RockYou password dataset~\cite{rockyou} contains the passwords of
around 32 million users of the RockYou gaming site.
The data-breach that produced the list was particularly costly as the
site did not bother hashing its users' passwords.
The list is complete, containing both very common passwords (the
password ``123456'' occurs 290729 times), as well as many unique ones
(2459760 passwords appear only once).
As a result, the list has been studied
extensively~\cite{bonneau2012guessing, bonneau2012science,
  weir2010testing, kelley2012guess,ur2012does}.
Fig.~\ref{fig:pass-dataset} (top) summarizes the frequency (dark
line) of the passwords in the whole dataset.
The passwords in the figure are ordered in decreasing frequency of appearance.

The dashed line in Fig.~\ref{fig:pass-dataset} (bottom) demonstrates
the strength of the passwords in the dataset using a simple metric
quantifying the likelihood of a successful brute-force attack against
a uniformly picked user in the dataset as a function of number of
guesses, assuming the attacker knew the exact distribution of
passwords in the dataset.
As a frame of reference, we also include a similar metric assuming the users picked their password uniformly from the 11884632
different passwords in the dataset (solid dark line) or if all 32
million users picked their passwords uniquely (solid light line).

As a candidate for the partitions, we group the passwords based on their 
frequency of appearance as an indirect measure of their cost. 
Namely, we group the passwords with the highest frequency in $\E_1$, passwords 
with the second highest frequency in $\E_2$, so on, which makes the last 
partition $\E_{2040}$ as the set of all the passwords that appear only once.
We use the inverse of frequency of a password as a rough estimate of
its usability cost.
After normalization, we set the usability costs to range from $C_1 = 
1/290729$ to $C_{2040} = 1$.
The cost associated with each partition can be seen as the dotted line
in Fig.~\ref{fig:pass-dataset} (top).

In Section~\ref{sec:example-capped-password-analysis} and
Section~\ref{sec:example-costly-password-analysis} we will compare the
behavior of users in the dataset to that of their equilibrium behavior
for each of our two attack settings.

\subsection{Cost Example: Cryptographic Keys}\label{sec:example-keys-model}

\begin{figure}
\includegraphics[width=0.95\linewidth]{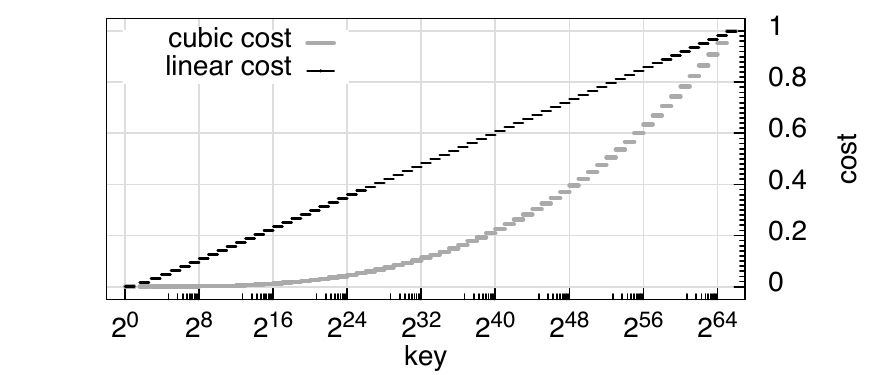}
\caption{\label{fig:keys-model} Synthetic cost model for selection of
  cryptographic keys, with cost proportional to linear and cubic power
  of key length.}
\end{figure}

The selection of secret keys for cryptographic protocols is usually
out of the hands of humans, nevertheless, the design decision of picking the
strength of the key (usually a function of its length) entails the same
cost/risk trade-offs.

Fig.~\ref{fig:keys-model} summarizes the space of possible keys for
two hypothetical cryptographic constructs.
In both we assume a key can be anywhere between 0 and 64 bits.
The examples differ only in the costs associated with each key.
A cost linear in the length of key approximately models the trade-off
in symmetric key systems such as AES which uses a number of rounds
proportional to key length. 
The cubic relation is more appropriate approximation of public/private
schemes such as naive implementations of RSA whose computation time
scales cubically with key length \cite{katz2014introduction}.
Note that in our analysis we will assume that the length of the key is
not known to a guesser, something that is usually not true of
public/private schemes.

In Sections \ref{sec:example-capped-keys-analysis},
\ref{sec:example-costly-keys-analysis},
\ref{sec:example-costly-keys-stack} we examine the equilibrium strategies of 
the picker
and guesser given each of these two cost models.

\section{Capped-Guesses}\label{sec:model:capped_guesses}
In the \emph{Capped-Guesses} scenario, the guesser, \emph{without observing the 
action of the picker}, chooses at most $K$ elements from the set of  
possible secrets $\mathcal{P}$, as his guesses. We assume, naturally, that $K$ 
does not depend on the actual guesses chosen.  The pure action of the guesser, 
which we denote by $A$, is  hence a subset of $\mathcal{P}$ of size $K$, since 
it is in the guesser's best interest to use all $K$ of his guesses. In 
the 
case of password selection, for example, each action represents an instance of a 
pre-computed table with which the guesser chooses to launch a dictionary attack. 
The action set of the guesser is therefore: $\mathcal{A}:=\{A| 
A\subset\mathcal{P}, |A|= K\}$,  the set of all possible pre-computed tables of 
size $K$.
The number $K$ represents the prowess of the guesser determined by the physical 
limitations in place: for instance, in the case of pre-computed table attacks on 
passwords, $K$ is determined by how much memory each hash entry occupies and how 
much total memory the attacker has available for the table. 
Alternatively, in an online password attack, it can be the number of tries he is 
allowed to make before getting locked out. 
We assume $K<|\Ps|=\sum_{i=1}^N|\E_i|$, since otherwise, the guesser trivially 
can find the secret with certainty. 
 A (pure) \emph{strategy profile} 
 here is simply a pair of picker and guesser actions, $(d,A)\in 
\mathcal{P}\times\mathcal{A}$.

The problem is the following: determine best strategies for the picker to 
choose her secret and the guesser to construct his guessing dictionary, when 
both parties are \emph{rational} decision makers. 
The problem can be modeled as a \emph{simultaneous move} game. Note that in 
game-theory, the term ``simultaneous move'' does not necessarily imply 
synchronicity, rather, the lack of observation of the move of other 
players (or any signal about it) before making a move. Otherwise, 
there is a sequentiality in the occurrences of the actions taken in our problem: 
the picker picks first. 
In our Capped-Guesses game, the ``actions'' and ``strategies'' simply coincide.
To complete the model, we next provide the utilities of the players given a 
strategy profile $(d,A)$.
Let $u_D$ and $u_A$ represent the utilities of the picker and the 
guesser respectively. Compactly put, we have:
\begin{align}
 u_D(d,A)=-c(d) -\lambda \mathbf{1}_{A}(d),& & &u_A(d,A)=\gamma 
\mathbf{1}_{A}(d)\label{eq:the_utilities}
\end{align}
where $\mathbf{1}$ represents the indicator function,\footnote{The indicator 
(characteristic) function of a subset $Y$ of a set $X$ is a function 
$\mathbf{1}_Y: X\to\{0,1\}$ defined as the following: $\mathbf{1}_Y(x)=1$ if 
$x\in Y$, and $\mathbf{1}_Y(x)=0$ if $x\notin Y$.} and 
$c(d):=\sum_{i=1}^NC_i\mathbf{1}_{\E_i}(d)$ is the usability cost of secret 
$d$. Note that this summation only contains one non-zero element: if the picked 
secret is from partition $\E_i$, she incurs the usability cost of $C_i$. 
A list of the main notations is provided in Table~\ref{Table:Notations}.

\begin{table}[ht]
\centering
\caption{List of main notations}\label{Table:Notations}
 \begin{tabular}{||c p{0.7\linewidth}||}\hline\hline
Notation & \quad Definition \\
\hline
$\mathcal{E}_i$ & {\footnotesize Set of secrets with the same cost of picking 
$C_i$}\\ 
$\mathcal{P},N$ & {\footnotesize Set of all secrets, Number of its partitions}\\
$C_i$ & {\footnotesize Cost of picking a secret from set $\mathcal{E}_i$, 
incurred by the picker}\\
$K$ & {\footnotesize Number of attempts available to the guesser in 
Capped-Guesses 
(size of his ``table'')}\\
$\sigma$ & {\footnotesize Cost of each attempt of the guesser in 
Costly-Guesses}\\
$\gamma$ & {\footnotesize Gain earned by the guesser if any of his guesses is 
correct \hide{(secret is found)}}\\
$\lambda$ & {\footnotesize Loss incurred by the picker if the secret is found 
{by the guesser} \hide{(secret is lost)}}\\ 
$c(d)$ & {\footnotesize Picker's usability cost for picking secret $d$. Short for $\sum_{i=1}^NC_i\mathbf{1}_{\E_i}(d)$.}\\ 
\hline
 \end{tabular}
\end{table}

\paragraph*{Solution Concept 1 -- Nash Equilibria (NE)}
A \emph{solution} of a game is a prediction of how rational players facing it 
may take decisions. A commonly used notion of a solution is  Nash Equilibrium 
(NE in short), which informally put, is a strategy profile that consists of 
simultaneously optimal responses to each other, keeping the others' strategies 
fixed, i.e., strategy profiles that are resistant against unilateral deviations 
of players.
Formally, in our two player game, this means the following:
the strategy pair $(d^*,A^*)\in \mathcal{P}\times\mathcal{A}$ is a (pure) NE if 
and only if: 
$d^*\in\arg\max_{d\in\mathcal{P}} u_{D}(d,A^*)$ and 
$A^*\in\arg\max_{A\in\mathcal{A}} u_A(d^*,A)$, 
in \eqref{eq:the_utilities}, 
i.e.,  $u_{D}(d^*,A^*)\geq u_{D}(d,A^*)$ for all $d\in\mathcal{P}$ and 
$u_A(d^*,A^*)\geq u_A(d^*,A)$ for all $A\in\mathcal{A}$.

It is not difficult to see that ``pure'' Nash Equilibrium is not a suitable 
solution 
concept for our game. In fact, except in trivial cases, no pure NE exists. This 
is because a pure strategy of the picker means selection of a specific secret. 
The 
best response of the guesser is then simply to include that secret in his guess 
dictionary. But then, the picker would have been better off to deviate and 
choose a  different secret and not incur the potentially huge cost of having her 
secret revealed with certainty.
In other words, deviation from any pure action (except in trivial cases) is 
beneficial for the picker. A similar argument can be made for the guesser. 

The above discussion motivates the search for a solution among \emph{mixed 
strategies}, which involve randomization thereby injecting ambiguity about the 
choice of each player. Specifically, a mixed strategy of a player is a 
probability distribution over her set of pure strategies.
For any finite nonempty set $\mathcal{S}$, let $\Delta(\mathcal{S})$ represent 
the set of all probability distributions over it. That is:
\[
\Delta(\mathcal{S}):=\{\vec\sigma\in\mathbb{R^+}^{|\mathcal{S}|}|\sum_{
s\in\mathcal{S}}\vec\sigma(s)=1\}
\]
For a given probability distribution 
$\vec\sigma\in\Delta(\mathcal{S})$, let the \emph{support} 
of $\vec\sigma$, or $\supp(\vec\sigma)$,  denote the subset of the domain of $\vec\sigma$ that 
receives a strictly positive probability, that is: 
$$\supp(\vec\sigma):=\{s\in\mathcal{S}|\vec\sigma(s)>0\}.$$
Moreover, for a given probability distribution 
$\vec\sigma\in\Delta(\mathcal{S})$ and a 
given subset $\mathcal{S}'\subseteq \mathcal{S}$, let $\vec\sigma(\mathcal{S}')$ 
represent the probability measure of $\mathcal{S'}$ with respect to 
$\vec\sigma$, that is, let 
$\vec\sigma(\mathcal{S'}):=\sum_{s\in\mathcal{S'}}\vec\sigma(s)$. 

Let $\vec\delta$ and $\vec\alpha$ represent a mixed strategy of the picker and 
guesser respectively. We hence have: 
$\vec\delta\in\Delta(\mathcal{P})$ and $\vec\alpha\in\Delta(\A)$.
Following a common abuse of notation in game theory, let  
$u_D(\vec\delta,\vec\alpha)$ and $u_A(\vec\delta,\vec\alpha)$ be the expected 
utility of the two players given a mixed strategy profile 
$(\vec\delta,\vec\alpha) \in \Delta(\mathcal{P})\times \Delta(\mathcal{A})$ 
where the expectation is taken with respect to the independent randomizations 
in the mixed strategies. That is: 
$u_D(\vec\delta,\vec\alpha):=\sum_{d\in\mathcal{P},A\in\A}u_D(d,
A)\vec{\delta}(d)\vec{\alpha}(A)$, and 
likewise for $u_A(\vec\delta,\vec\alpha)$. Replacing 
from~\eqref{eq:the_utilities}, we have:
\begin{subequations}\label{eq:exptec_utilities_mixed}
\begin{align}
&u_D(\vec\delta,\vec\alpha)\!=\!-\!\sum_{d\in\mathcal{P}}c(d)\vec\delta(d) 
\!-\!\lambda \!\!\!\!\!\sum_{d\in\mathcal{P},A\in\A}\!\!\!\!\!
\mathbf{1}_{A}(d)\vec\delta(d)\vec\alpha(A)\label{
subeq:expected_utility_mixed_defender}\\
&u_A(\vec\delta,\vec\alpha)=\gamma\!\!\!\sum_{d\in\mathcal{P},A\in\A}\!\!\!
\mathbf{1}_{A}(d)\vec\delta(d)\vec\alpha(A)\label{
subeq:expected_utility_mixed_attacker}
\end{align}
\end{subequations}
Note that we are assuming randomization  {per se} is costless.
A mixed strategy of the guesser, $\vec\alpha\in\Delta(\mathcal{A})$, specifies 
the probability that each feasible dictionary (table) is selected. For our 
model, it is often simpler to instead specify the marginal probabilities that 
each secret is tried by the guesser. Specifically, let us define $\vec{\rho}$ 
such that $\vec\rho(d)$ denotes the probability that secret $d$ is in the 
($K$-sized) table of the guesser. $\vec{\rho}$ and $\vec\alpha$ are related 
through: 
$\vec\rho(d)=\sum_{A\in\A} \mathbf{1}_{A}(d)\vec\alpha(A)$.
Moreover, using the notion of probability measure and the fact that all members 
of the same partition by definition have the same choosing cost for the picker, 
we have: $\sum_{d\in\Ps} c(d)\vec\delta(d)=\sum_{i=1}^N C_i\vec\delta(\E_i)$. 
Hence, the expressions in \eqref{eq:exptec_utilities_mixed} can be simplified 
as:
\begin{equation}
\begin{aligned}\label{eq:exptec_utilities_mixed_rewritten}
u_D(\vec\delta,\vec\rho)=&-\sum_{i=1}^N C_i\vec\delta(\E_i) -\lambda  
\sum_{d\in\mathcal{P}} \vec\delta(d)\vec\rho(d), 
\\ 
u_A(\vec\delta,\vec\rho)=&\gamma\sum_{d\in\mathcal{P}}\vec\delta(d)\vec\rho(d) 
\end{aligned}
\end{equation}

A mixed NE is defined in the same way as a pure NE, except that the optimization 
variables and the optimization spaces are replaced accordingly. The set of pure 
NE are contained in the set of mixed NE, since pure strategies can be obtained 
from degenerate distributions over the strategies. That is, a mixed strategy 
profile $(\vec{\delta^*},\vec{\alpha^*})$ is a mixed NE iff:
\begin{gather*}
 u_{D}(\vec{\delta^*},\vec{\alpha^*})\!\!\!\underset{\forall 
\vec{\delta}\in\Delta\mathcal{P}}{\geq}\!\!\! 
u_{D}(\vec{\delta},\vec{\alpha^*}),\ \ u_A(\vec{\delta^*},\vec{\alpha^*}
)\!\!\!\underset{\forall \vec{\alpha}\in\Delta\mathcal{A}}{\geq} 
\!\!\!u_A(\vec{\delta^*},\vec{\alpha}). 
\end{gather*}

\paragraph*{Solution Concepts 2 \& 3 --  maximin and minimax}
A (mixed) strategy of the picker $\vec\delta^{\mathrm{maximin}}\in\Delta(\Ps)$ 
is a \emph{maximin} strategy of hers if and only if:
\begin{gather*}
\vec\delta^{\mathrm{maximin}}\in\arg\max_{\vec\delta\in\Delta(\Ps)}[\min_{
\vec\alpha\in\Delta(\A)}u_D(\vec\delta,\vec\alpha)]
\end{gather*}
Let 
$\underline{u_D}(\vec\delta):=\min_{\vec\alpha\in\Delta(\A)}u_D(\vec\delta,
\vec\alpha)$, which is the worst utility of the picker among all reactions 
of the guesser if she chooses the mixed strategy of $\vec\delta$. Then 
$\vec\delta^{\mathrm{maximin}}$ maximizes $\underline{u_D}(\vec\delta)$, 
achieving the \emph{maximin utility}, which we will denote by $\underline{u_D}^{\max}$.
This is the mixed strategy that guarantees (secures) the picker at least her 
maximin utility irrespective of the strategy of the guesser. For this reason, 
maximin strategies are sometimes also referred to as ``security'' strategies. 
maximin strategies are 
recipe for action when a player is 
strategically pessimistic, in that she believes the opponent(s) behave in 
such a way to  hurt her utility the most, as opposed to selfishly maximize 
their own utilities. Hence, the focus is solely on the utility of that player, 
and rationality of other players is not taken into account.

This is conceptually different from a \emph{minimax} strategy of a player. 
Formally, 
$\vec\delta^{\mathrm{minimax}}$ is a picker's minimax strategy if and only if:
$\vec\delta^{\mathrm{minimax}}\in\arg\min_{\vec\delta\in\Delta(\Ps)}[\max_{
\vec\alpha\in\Delta(\A)}u_A(\vec\delta,\vec\alpha)]$.
Let $\overline{u_A}(\vec\delta):=\max_{\vec\alpha\in\Delta(\A)}u_A(\vec\delta,
\vec\alpha)$, which is the best utility of the guesser  among all of his reactions if the picker chooses the mixed strategy of 
$\vec\delta$. Then  $\vec\delta^{\mathrm{minimax}}$ minimizes 
$\overline{u_A}(\vec\delta)$, 
guaranteeing that the utility of the guesser is bounded by his \emph{minimax 
utility}, denoted by 
$\overline{u_A}_{\min}$. 
That is the strategy that the picker can adopt to hurt the utility of the 
opponent (the guesser) the most, ignoring her own utility. In zero-sum games, 
the utility of each player is negative (i.e., additive inverse) of the of 
other. Hence, hurting the expected pay-off of the opponent the most is 
exactly 
equivalent to helping your own expected pay-off the most. This means that 
minimax and maximin strategies of each of the players coincide.
But this in 
general does \emph{not} extend  to non-zero-sum 
games. This is exactly the situation in our game. It is easy to see that the 
minimax strategy of the picker is simply to uniformly randomize over the 
\emph{entire} set of secrets, effectively maximizing the ambiguity, minimizing 
any useful information that the guesser can exploit. However, this completely 
ignores the cost of choosing costly secrets. As we will show, the maximin 
strategy of the picker is in general different from uniform randomization over 
the entire set of secrets.

Likewise, we can speak of the maximin and minimax strategies of the guesser: 
$\vec\alpha^{\mathrm{maximin}}\in\Delta(A)$ is a maximin strategy of the 
guesser 
if and only if:
$\vec\alpha^{\mathrm{maximin}}\in 
\arg\max_{\vec\alpha\in\Delta(\A)}\underline{u_A}(\vec\alpha)$ where 
$\underline{u_A}(\vec\alpha):=\min_{\vec\delta\in\Delta(\Ps)}u_A(\vec\delta,
\vec\alpha)$.
Here also the distinction between the maximin and minimax strategies can be 
observed.
Specifically, if the guesser is on the (pessimistic) belief that the picker is 
trying to hurt his utility the most (or equivalently plan according to the 
``worst case scenario'' of the strategy of the picker irrespective of her 
rationality), he should select his $K$ guesses uniformly randomly over the 
entire set of secrets.
This approach  ignores the pay-off structure of the picker and hence does not 
take advantage of the presence of the preferences of the picker over the 
secrets. We will see how the guesser can exploit this knowledge in 
\Section~\ref{sec:Analysis:capped_guesses}.

\paragraph*{Solution Concept 4 --  Strong Stackelberg Equilibria (SSE)}
Consider the situation in which the picker has the power of credible commitment 
to a mixed strategy. Note that this is in general different from commitment to a 
pure strategy and requires a different ``apparatus''.
The relevant solution concept for these cases is the Strong Stackelberg 
Equilibria, which intuitively put, are the best mixed strategies that the leader 
(picker in our case) can commit to, knowing that the follower (guesser, here) 
will observe this commitment and will respond selfishly optimally to it. 
In order for the solution concept to exist, it also needs the extra assumption 
that whenever the follower is indifferent between a set of best responses, he 
will break ties in favor of the leader. This is a benign assumption, because the 
leader can turn any of the indifferent best responses of the follower to a 
strict preference through an infinitesimal modification of her mixed strategy. 
Note that a (pure) strategy of the follower is now a function of the commitment 
distribution of the leader. That is, if the follower is the guesser, a pure 
strategy of the follower is a mapping from $\Delta(\Ps)$ to $\A$.
Formally, $(\vec{\delta^*},\vec\alpha^{\mathrm{BR}})$ in which 
$\vec{\delta^*}\in\Delta(\Ps)$ and 
$\vec\alpha^{\mathrm{BR}}:\Delta(\Ps)\to\Delta(\A)$, constitutes a SSE if and 
only if:\footnote{The superscript $\mathrm{BR}$ is chosen to stand for ``best 
response''.}
\begin{enumerate}
 \item $ \vec{\delta^*}\in\arg\max_{\vec\delta\in\Delta(\Ps)} 
u_{D}(\vec\delta,\vec\alpha^{\mathrm{BR}}(\vec\delta))$
 \item $ 
\vec\alpha^{\mathrm{BR}}(\vec\delta)\in\arg\max_{\vec\alpha\in\Delta(\A)}
u_A(\vec\delta,\vec\alpha)$ 
 \item $ 
\vec\alpha^{\mathrm{BR}}(\vec\delta)\in\arg\max_{\vec\alpha'\in\arg\max_{
\vec\alpha\in\Delta(\A)}u_A(\vec\delta,\vec\alpha)}u_{D}(\vec\delta,
\vec\alpha')$ 
\end{enumerate}

\begin{figure}
\begin{framed}
\begin{center}
\textbf{Game 1:} {Capped-Guesses} 
\end{center}
\begin{description}
	\item[Players:] \textsc{\texttt{Picker, Guesser}}
	\item[Strategy Sets:] \textsc{\texttt{Picker's}}: $\{d\in\Ps\}$\\ 
	\textsc{\texttt{Guesser's}}: $\{A\subset \Ps,\ |A|= K\}$ 
	\item[Utilities:]  \textsc{\texttt{Picker}}: $u_D(d,A)=-c(d) -\lambda 
\mathbf{1}_{A}(d)$,\\
	\textsc{\texttt{Guesser}}: $u_A(d,A)=\gamma \mathbf{1}_{A}(d)$%
\end{description}
\end{framed}
\label{fig:Game1}
\end{figure}

\section{Analysis of the Capped-Guesses 
Scenario}\label{sec:Analysis:capped_guesses}
As our main result for the Capped-Guesses scenario, we provide a sufficient 
condition for a strategy pair to be a mixed NE (Prop.~\ref{Prop:NE}). 
We show that the NE and 
maximin strategies of the picker coincide 
(Lemma~\ref{Lem:NE_is_Maximin}).
This useful property leads us to other implication: all NE are 
interchangeable 
(Corollary~\ref{Corr:Interchangeability_I}) and they all yield the same utility 
for the picker (Corollary~\ref{Corr:Interchangeability_II}). Another 
implication of the lemma is that 
for this scenario, the set of optimal mixed strategies of the picker to commit 
to, i.e., her SSE strategies, are also the same as her NE strategies, and 
moreover, they attain her the same utility as any NE does 
(Corollary~\ref{Corr:SSE_NE_equivalence}). Finally, we provide a mild 
constraint under which the sufficient conditions provided in 
Prop.~\ref{Prop:NE} for a mixed strategy of the  picker to be a NE are 
also necessary conditions, implying uniqueness of the description of the NE for 
almost all instances of the game (Corollary~\ref{Corr:uniqueness}). 
These results fully characterize the solution of the Capped-Guesses game.
The proofs of the results in this section can be found in 
\ifcameraready
\else Appendices
\ref{Appendix:Proof_of_Prop:NE} through
\ref{Appendix:Proof_of_Corr:Uniqueness}. \fi

First, note that following Nash's Theorem, our finite  game has at least one 
mixed NE. The existence of maximin, 
minimax and SSE solutions also follow standard results in game theory 
\cite{basar1995dynamic}. In order to explicitly describe the NE, we need to 
define a few parameters. Let:
$L:=\min_{1\leq l\leq N} l\ \mathrm{s.t.}\ \sum_{i=1}^l|\E_i|>K$.
Note that in part this means: $\left|\cup_{i=1}^m \E_i\right|\leq K$ for
any $m < L$ (recall 
that $K$ is the dictionary size of the guesser -- the available number of 
guesses to the adversary). 
Now suppose the picker chooses her secret according to a randomization only 
from the first $m$ (cheapest) partitions where $m<L$. Then the guesser can 
correctly guess the secret with certainty, because he can simply include the 
entire $\cup_{i=1}^m \E_i$ in his guessing dictionary. Hence, for the 
picker, the (strictly) best among such options that lead to certain loss of the 
secret is simply picking from the cheapest partition which yield her a utility 
of $-C_1-\lambda$.\footnote{In the language of game theory, any mixed strategy 
of the picker that only randomizes over $\cup_{i=1}^m \E_i$ where $m<L$ is 
strictly dominated by strategies that only randomize over $\E_1$.} 
The picker can reduce the  chance of 
a correct guess by randomizing over partitions beyond $\cup_{i=1}^{m} 
\E_i$, but then the picker has to balance usability costs with the gain in 
increasing the entropy. Define: 
\begin{gather}
\mathcal{J}\!:= \!\left\{\!L< j\leq N\mid {\lambda 
K+\sum_{i=1}^{j-1}C_i|\E_i|}\geq 
C_j{\sum_{i=1}^{j-1}|\E_i|}\right\}.\label{eq:J_Definition} 
\end{gather}
That is, $\mathcal{J}$ characterizes the partitions for which the inequality 
of $\lambda K/(\sum_{i=1}^{j-1}|\E_i|) 
+ (\sum_{i=1}^{j-1}C_i|\E_i|)/(\sum_{i=1}^{j-1}|\E_i|)  \geq C_j
$ holds. Since only $j> L$ are considered, we have 
$K<(\sum_{i=1}^{j-1}|\E_i|)$. In particular, suppose the picker \emph{uniformly}
randomizes over $\unif(\cup_{i=1}^{j-1} \E_i)$. Then, irrespective of the 
strategy of the guesser as long as its support is $\unif(\cup_{i=1}^{j-1} 
\E_i)$, his chance of finding the secret is exactly 
$K/(\sum_{i=1}^{j-1}|\E_i|)$, and hence the security cost of the picker is 
$\lambda K/(\sum_{i=1}^{j-1}|\E_i|)$. 
Moreover, the usability cost of the picker for uniformly randomizing over 
$\unif(\cup_{i=1}^{j-1} \E_i)$ is 
$(\sum_{i=1}^{j-1}C_i|\E_i|)/(\sum_{i=1}^{j-1}|\E_i|)$. 
Therefore, the condition in the definition of $\mathcal{J}$ translates to the 
following: $j\in \mathcal{J}$ if the usability cost of choosing from $\E_j$ is 
less than the overall cost (security and usability cost) of uniformly 
randomizing over the (combined) first $j-1$ (cheapest) 
partitions.\footnote{With simple algebra, the condition can be shown to be 
equivalent to the following:  $\big[\lambda K/(\sum_{i=1}^{j-1}|\E_i|) 
+ (\sum_{i=1}^{j-1}C_i|\E_i|)/(\sum_{i=1}^{j-1}|\E_i|)  \big] \geq
\big[\lambda K/(\sum_{i=1}^{j}|\E_i|) 
+ (\sum_{i=1}^{j}C_i|\E_i|)/(\sum_{i=1}^{j}|\E_i|)  \big]$. In 
words, $j\in\mathcal{J}$ if the overall cost of 
uniformly randomizing over the combined first $j-1$ partitions is  
more than that of uniformly randomizing over the combined first $j$ partitions 
for the picker. This in turn implies that, for the picker, uniform 
randomization over the first $j-1$ partitions is (weakly) \emph{dominated} by 
uniformly randomizing over the first $j$ partitions.}

If $\mathcal{J}\neq\emptyset$, define $J:=\max \mathcal{J}$. We label the cases 
where either $\mathcal{J}=\emptyset$ or $C_1+\lambda\leq 
\left({\sum_{i=1}^JC_i|\E_i|+\lambda 
K}\right)/\left({\sum_{i=1}^J|\E_i|}\right)$ as 
``\emph{total defeat}'', since in such cases the picker chooses her 
secret from the cheapest partition, $\E_1$, knowing that her choice will be 
guessed correctly, because it is not worthwhile (or not possible) for her to try to 
prevent it. 
We will refer to all other situations, i.e., when we have 
$\mathcal{J}\neq\emptyset$ \emph{and} 
$C_1+\lambda>\left({\sum_{i=1}^JC_i|\E_i|+\lambda 
K}\right)/\left({\sum_{i=1}^J|\E_i|}\right)$ as 
``\emph{ordinary}'' cases, since, as we show, it is worthwhile for the picker to try 
to avoid certain revelation of her secret. 

Recall that 
$\vec{\rho^*}(p)=\sum_{A\in\A}\vec{\alpha^*}(A)\mathbf{1}_{A}(p)$ is just 
the probability that secret $p$ will be among the $K$ selections of the 
guesser, 
given his mixed strategy of $\alpha^*$.
We now mathematically present the NE strategies and subsequently describe them 
in words: 
\begin{proposition}\label{Prop:NE}
For the ``ordinary'' cases in a Capped-Guesses game,  consider a strategy pair 
$(\vec{\delta^*},\vec{\alpha^*})$ 
where:
\begin{gather*}
  \vec{\delta^*}=\unif(\cup_{i=1}^J \E_i),
\end{gather*}
and:
\begin{subequations}
\begin{align}
&\vec{\rho^*}(p)=\frac{K}{\sum_{j=1}^J|\E_j|}+B_i, \ \ \text{$\forall p\in\E_i$ 
where $i\leq J$} \label{eq:Alpha^*_below_J}\\
&\vec{\rho^*}(p)=0, \ \ \text{$\forall p\in\E_i$ where $i> 
J$}\label{eq:Alpha^*_beyond_J}
\end{align}\label{eq:CaseN_alpha*}
\end{subequations}
where 
$B_i\!:=\!\!\left.\left[{\sum_{j=1}^{J}\!C_j|\E_j|\!-\!C_i\!\sum_{j=1}^{J}
\!|\E_j| } \right ]\!
\middle/\!\left[{\lambda  \sum_{j=1}^J\!|\E_j|}\right]\right.$. Then, the 
strategy 
pair $(\vec{\delta^*},\vec{\alpha^*})$ is a (mixed) NE.
For the ``total defeat'' cases, consider a strategy pair 
$(\vec{\delta^*},\vec{\alpha^*})$ that satisfies the following:
\begin{align}
\text{Picker: }&\vec{\delta^*}(\E_1)=1\\ 
  \text{Guesser: }&\begin{cases} \vec{\rho^*}(p)> 1-\frac{C_i-C_1}{\lambda} & 
\forall p\in\E_i, i\leq J, \\
  \vec{\rho^*}(p)=0 & \forall p\in\E_i, i> J 
\label{eq:CaseN_alpha*_total_defeat}
  \end{cases}
\end{align}
Then $(\vec{\delta^*},\vec{\alpha^*})$ constitutes a NE.
\end{proposition}

In words, for the ``ordinary'' cases, the proposed NE is the following: the 
picker chooses her secret \emph{only} 
from the first $J$ partitions, i.e., the $J$ most favored partitions, and does 
so \emph{uniformly} randomly.  Note in particular that the preference profile 
of the picker only affects her NE strategy through the number of partitions that 
constitute the domain of secrets to choose from, but the randomization over 
this domain is always uniform, despite the uneven preferences over them. 

On the other hand, the guesser, knowing the picker does not choose her secret 
with any positive 
probability from partitions beyond 
$\E_J$, does not include any guesses from them either 
\eqref{eq:Alpha^*_beyond_J}. 
The guesser selects uniformly randomly \emph{within} partitions $1,\ldots,J$ 
but \emph{not across} them. That is, even though the secrets from the same 
partition are equally likely to be part of the guessing dictionary of the 
guesser, the secrets from partition $i\leq J$ are chosen with a \emph{bias} 
equal to $B_i$ away from uniform guessing. This is despite the fact that 
the picker chooses her secret uniformly randomly from the first $J$ partitions. 
Indeed, as we discuss 
in the proof, the guesser explores the relatively favored partitions of the 
picker among the first $J$ partitions with a positive bias compared to her 
relatively less favored partitions. Specifically, the bias is exactly 
such that the picker is indifferent about choosing the secret from any of the 
first $J$ partitions. 

For the cases of ``total defeat'', the picker simply chooses her secret from 
partition $\E_1$, the least costly partition, 
and the guesser includes all of that partition into his dictionary, along with 
other partitions such that the picker is forced into picking her secret only 
from the cheapest partition.   Thus, the secret will be discovered by the 
guesser with probability one. Note that, interestingly, the NE was not 
at all affected by $\gamma$, the gain parameter of the guesser.

Our next series of results describe the properties of the NE in regards to 
other strategic metrics. Note that establishing these results do not rely 
directly on the explicit expression of the NE in Prop.~\ref{Prop:NE}.

In general, playing NE strategies by by a player conjures the assumption that 
the other player(s) are 
indeed \emph{rational}, in that, they are interested in maximizing their own 
utility as opposed to antagonistically trying to minimize the utility of that player. 
But what if this rationality assumption cannot be made in our case regarding 
the guesser?
Our next observation dispels that concern by establishing that for the 
Capped-Guess scenarios, 
NE strategies of the picker are her maximin strategies and vice versa. 
\begin{lemma}\label{Lem:NE_is_Maximin}
Let $\Omega^{\mathrm{D}}_{\mathrm{NE}}$ be the set of NE strategies of the 
picker and
 $\Omega^{\mathrm{D}}_{\mathrm{maximin}}$ be the set of her maximin strategies 
in a game of Capped-Guesses.
We have: 
$\Omega^{\mathrm{D}}_{\mathrm{NE}}=\Omega^{\mathrm{D}}_{\mathrm{maximin}}$.
\end{lemma}

The lemma establishes that the picker can randomize 
according to her NE and (in expectation) be guaranteed at
least the expected utility prescribed by the NE, irrespective of the mixed 
strategy of the guesser, be it a NE or not. From a different viewpoint, the 
picker can act 
according to her pessimistic 
maximin strategy, but be assured that she does not lose anything in expectation 
by not playing a NE. Note that this property only holds for the NE strategy of 
the picker and not of the guesser (Recall that the maximin strategy of the 
picker is choosing his $K$ guesses uniformly randomly from the entire 
secret space $\Ps$). 

Here, we just mention the gist of the proof.\ifcameraready
\else Detail of the proof is in Appendix~\ref{Appendix:Proof_of_Prop:NE}.\fi
The argument starts by noting from
\eqref{eq:exptec_utilities_mixed} that for any $\vec\delta\in\Delta(\Ps)$, 
$\vec{\alpha^*}\in\arg\max_{\vec\alpha\in\Delta(\A)} 
u_{A}(\vec\delta,\vec\alpha)$ if and only if:  
$\vec{\alpha^*}\in\arg\min_{\vec\alpha\in\Delta(\A)} 
u_{D}(\vec\delta,\vec\alpha)$.
To see this, note that the pay-off of the picker is 
composed of two parts, the first part is the expected cost of choosing the 
secret, and the second part is the expected cost of losing it. For any given 
mixed strategy of the picker, the guesser can only affect the second part of 
the utility of the picker. Specifically, 
$u_{D}(\vec\delta,\vec\alpha)=-(\lambda/\gamma) 
u_{A}(\vec\delta,\vec\alpha)+\phi(\vec\delta)$, where $-(\lambda/\gamma)<0$ and 
$\phi(\vec\delta)$ is an expression that does not depend on $\vec\alpha$. 
That 
is, the (rational) best response of the guesser to any ``given''  
strategy of the picker, also yields the worst utility for the picker. Hence, 
assuming a rational best response and strategically worst case scenario become 
equivalent for the picker.

Next two results (corollaries of Lemma~\ref{Lem:NE_is_Maximin}) establish the 
interchangeability of the NE and remove the concern of ``Equilibrium Selection'' 
in games of Capped-Guesses.
\begin{corollary}\label{Corr:Interchangeability_I}
 \textbf{Interchangeability of NE (I)}: If 
$(\vec{\delta^*}_1,\vec{\alpha^*}_1)$ 
and $(\vec{\delta^*}_2,\vec{\alpha^*}_2)$ 
are both NE in a game of Capped-Guesses, then so are 
 $(\vec{\delta^*}_1,\vec{\alpha^*}_2)$ and 
$(\vec{\delta^*}_2,\vec{\alpha^*}_1)$.
\end{corollary}

This corollary shows that if at all there are more than one distinct NE 
present, then no matter which NE strategy each player chooses to play, the outcome is still a 
NE. The next corollary further shows that, even if there were multiple NE, 
there is no question of preference between them for the picker, since her 
utility is the same in all of them: 
\begin{corollary}\label{Corr:Interchangeability_II}
 \textbf{Interchangeability of NE (II)}: All NE in a Capped-Guesses game yield 
the same utility for the picker. 
Specifically, if $(\vec{\delta^*},\vec{\alpha^*})$ is a NE of the 
Capped-Guesses 
game, then: 
 $u_D(\vec{\delta^*},\vec{\alpha^*})=\underline{u_D}^{\max}$. 
 \end{corollary}

These two results imply that, as far as the picker is concerned, it suffices to 
to find ``a'' NE, as we did in Prop.~\eqref{Prop:NE}, which is in general 
easier that finding 
the set of all NE. Although in our game, we will show that, almost in all 
cases, the NE is in fact unique (Corollary~\ref{Corr:uniqueness}).
The next corollary states that a NE strategy of the picker is also an optimum 
strategy of her to commit to, and  {vice versa}. 
\begin{corollary}\label{Corr:SSE_NE_equivalence}
In a Capped-Guesses game, let  $\Omega^{\mathrm{D}}_{\mathrm{SSE}}$ be the set 
of picker's SSE strategies. Then: 
$\Omega^{\mathrm{D}}_{\mathrm{SSE}}=\Omega^{\mathrm{D}}_{\mathrm{NE}}$.
\end{corollary}

As in Lemma~\ref{Lem:NE_is_Maximin}, the corollary follows by showing that given 
the committed strategy of 
the picker, the guesser will try to maximize his own utility, which in our 
Capped-Guesses game, is exactly what he would do if he wanted to minimize the 
utility of the picker. Hence the best mixed strategy to commit to by the picker 
is exactly the strategy that maximizes her minimum utility, i.e., her maximin 
strategy, which we previously showed to match the NE strategies. 
Intuitively, this is because in the Capped-Guesses model, the guesser will 
enter the game irrespective of the randomization strategy of the picker, 
and use all of his $K$ attempts. Moreover, he chooses his $K$ guesses so as to 
maximize the chances of finding the secret, which is exactly antagonistic to 
the utility of the picker given the randomized strategy of the picker (refer 
to the discussion after Lemma~\ref{Lem:NE_is_Maximin}). 

Note that the ability to commit to a mixed strategy is guaranteed not to hurt 
the ``committer'' (leader), since the leader can always commit to her Nash 
strategies
and yield at least her Nash utilities \cite{von2010leadership}. Or commit to a 
maximin strategy and guarantee her maximin utility. But in general, she 
may 
be able to do better and improve upon her Nash Equilibria. Even in the presence 
of Corollary~\ref{Corr:SSE_NE_equivalence}, due to the property that in SSE, 
the 
follower breaks ties among his best responses in favor of the leader, identical 
SSE and NE strategies of the picker may lead to distinct utilities for her. 
However, the  following lemma establishes that for the game of Capped-Guesses, 
this is not the case: the power to commit does not ``buy'' the picker any extra 
benefit. Specifically, the utility of the picker when best-committing is no 
better than 
her maximin utility. 
\begin{corollary}\label{Corr:no_commitment_value}
Let  $(\vec{\delta^*},\vec\alpha^{\BR})$ be a SSE of a Capped-Guesses game. 
Then 
we have: 
$u_D(\vec{\delta^*},\vec\alpha^{\BR}(\vec{\delta^*}))=\underline{u_D}^{\max}$. 
\end{corollary}

This result can also be expressed in the measure of the ``value of mixed 
commitment'' as discussed in \cite{letchford2012value}: the value of mixed 
commitment for the picker in Capped-Guesses games is one, i.e., commitment 
achieves nothing above what is achievable in NE, and hence there is no 
advantage in 
commitment. As we will see in Section~\ref{sec:Analysis:costly_guesses}, this 
is \emph{drastically} different from the situation in
Costly-Guesses scenarios.

\begin{corollary}\label{Corr:NE_SSE_MAximin_Value}
 The NE strategies of the picker as described in Prop.~\ref{Prop:NE} are 
also maximin and SSE.
Moreover, her utility in all NE and SSE is her maximin utility,  given 
as: $\underline{u_{D}}^{\max}\!=\! -\left.\left[\sum_{j=1}^{J}C_j|\E_j|+\lambda 
K\right]\middle/\left[{ \sum_{j=1}^J|\E_j|}\right]\right.$ in ``ordinary'' 
cases, and  
 $\underline{u_{D}}^{\max}\!=\!-C_1-\lambda$ in ``total defeat'' cases.
\end{corollary}

The next corollary is rather less important in characterization of this game in 
the light of Corollary~\ref{Corr:Interchangeability_II}. 
Nevertheless, it also shows that not only the utilities, but in fact even the 
equilibrium strategies themselves are almost always unique. 
This removes the question whether there may be other simpler to play NE of the 
game than presented in Prop.~\ref{Prop:NE} (even though the NE for the 
picker is quite simple as is). The answer is no, almost never.  
Referring to the definition of $J$ in \eqref{eq:J_Definition}, it allows to 
have 
 $\left(\sum_{j=1}^{J-1}C_j|\E_j|+\lambda 
K\right)=C_J\left(\sum_{j=1}^{J-1}|\E_j|\right)$.
We will refer to such a case as a \emph{degenerate case}, which is completely 
identifiable from the parameters of the problem. For all other 
(``non-degenerate'') cases, the condition 
$\left(\sum_{j=1}^{J-1}C_j|\E_j|+\lambda 
K\right)\geq C_J\left(\sum_{j=1}^{J-1}|\E_j|\right)$ is \emph{strictly} 
satisfied.  

\begin{corollary}\label{Corr:uniqueness}
 Aside from degenerate cases identified above, the sufficient conditions 
for 
a NE strategy profile provided in Prop.~\ref{Prop:NE} are also necessary. 
\end{corollary}

Note that the corollary in part implies that for ``non-degenerate'' ``ordinary'' 
cases, the NE is unique.

\subsection{Equilibrium Example: 
Passwords}\label{sec:example-capped-password-analysis}

\begin{figure}
\includegraphics[width=0.95\linewidth]{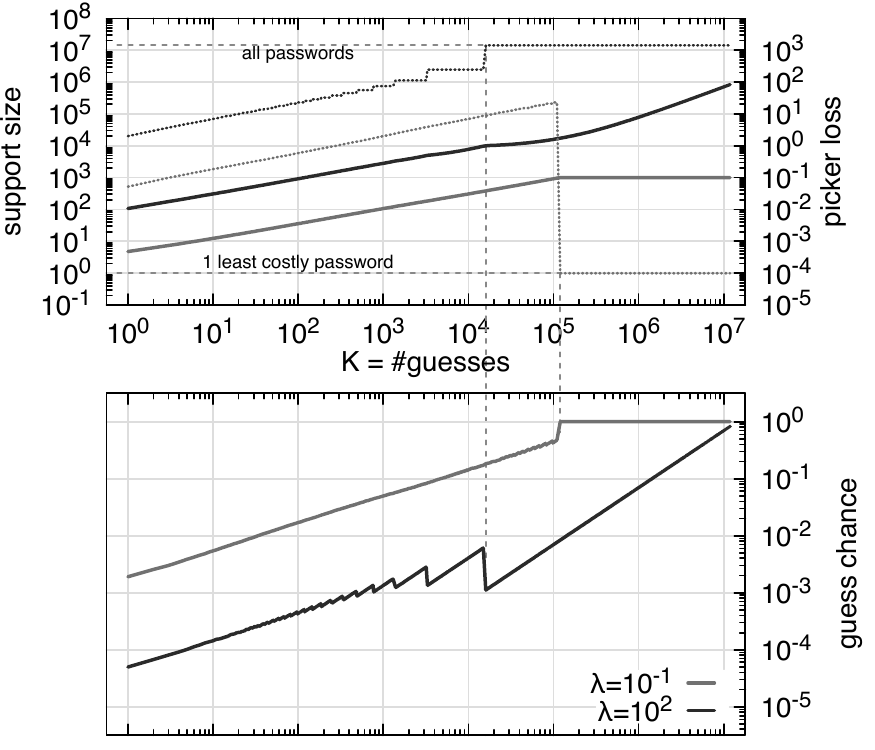}
\caption{\label{fig:pass-capped-model} (Top) Picker loss (solid) and
  corresponding key space size (dotted) as a function of the number of
   available guesses to the adversary for different values of loss ($\lambda$).
  (Bottom) Chance of a correct password guess as a function of the number of 
guesses
  in RockYou dataset and for a range of loss ($\lambda$) values.}
\end{figure}

Fig.~\ref{fig:pass-capped-model} summarizes the equilibrium
behaviors of the picker engaged in the Capped-Guesses game.
The top part of the figure shows picker loss (negative of the utility)
as solid lines and the size of the support set over which the picker
chooses his passwords, as the dotted lines.
The bottom part of the figure shows the probability that the password
would be found by the guesser.
All of these are shown as functions of the number of guesses available to
the guesser and for two different values of $\lambda$.
Recall that the cost of picking passwords was normalized. 
In this manner $\lambda$ serves as the cost of security losses for having the
password guessed relative to the usability cost. 

As the adversary is granted more guesses, the picker has to include a 
larger subset of passwords to (uniformly) randomize over.
When this support set is exhausted or the additional cost exceeds
the benefits, the picker gives up and picks only the cheapest password (``total 
defeat'').
For high values of $\lambda$, the picker never gives up, specifically, for large enough number of the available guesses, the picker uniformly randomizes over the 
entire set of passwords. 

\hide{As a point of comparison, the bottom portion of
Fig.~\ref{fig:pass-capped-model} also includes the success chances of the 
guesser when the picker is known to pick passwords like the average
user from the dataset (dashed line).
Though our model is a simplification of the preferences of users and
their assessments of risk, if we were to interpret the user as one
behaving rationally in our model, then the average RockYou user
valuates the cost of losing his or her password at around 10\% of the
cost of picking the most difficult password.}

\subsection{Equilibrium Example: Cryptographic
  Keys}\label{sec:example-capped-keys-analysis}

\begin{figure}
  \includegraphics[width=0.95\linewidth]{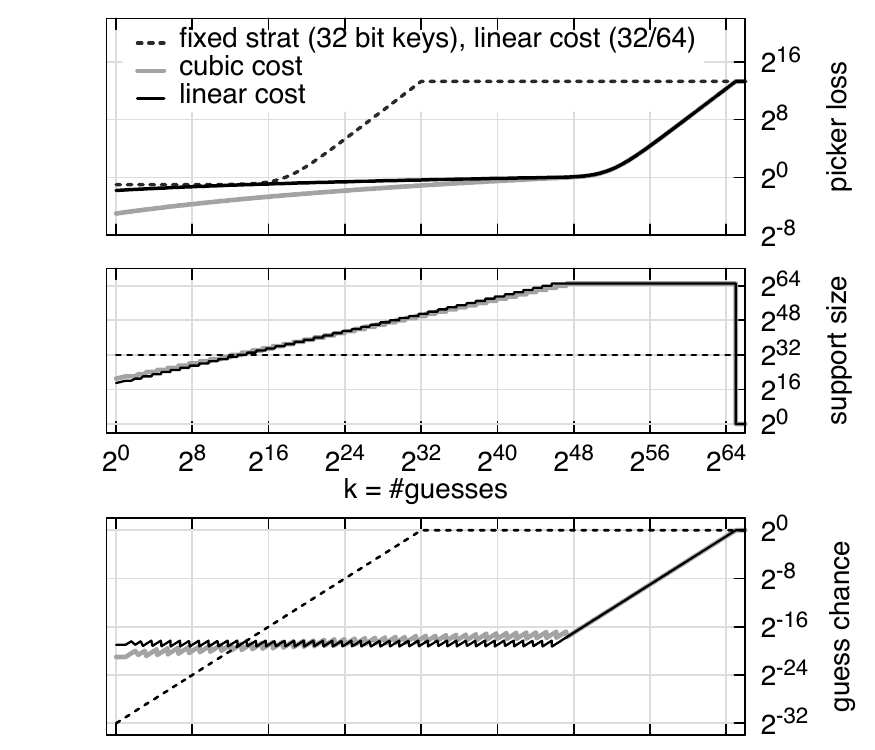}
  \caption{\label{fig:keys-capped-model} Equilibrium in the Capped-Guesses 
model 
for key selection.
    (Top) picker loss as a function of number of available guesses.
    (Middle) the corresponding size of the support set.
    (Bottom) the resulting chance of picker successfully finding the key.
    All with $\lambda=1000$.}
\end{figure}

Fig.~\ref{fig:keys-capped-model} demonstrates the result of
equilibrium key picking and key guessing.
The top part of the figure shows the loss incurred by the picker in
the two different cost models (linear or cubic in key size) as a
function of the number of available guesses.
The middle part of the figure focuses on the size of the key space that the
picker is forced to choose from.
The bottom of the figure shows the resulting probability that the key
will be discovered using $K$ brute-force guesses.
For brevity we only included in the graph the results for when $\lambda = 1000$.
Note that the differing  cost models do not have much impact
in this scenario and are overshadowed by the magnitude of the
exponentially increasing key space that is available to the picker.

For comparison purposes, the dashed line in the figure represents the
fixed picker behavior that chooses from among only the 32 bit keys and incurring a linear cost of picking, 0.5.
The top of the figure shows that this strategy does worse in terms of
loss than one which responds to adversary's power.

\section{Costly-Guesses}\label{sec:model:costly_guesses}
An alternative setting to a guesser with a limited number of guesses
is one with costly actions: consider a guesser that incurs a cost of
$\sigma>0$ per each guess.
To keep the model and analysis simple, we assume that this cost is not
guess-dependent.
In the case of passwords, for instance, this means that computation of
the hash of a guess is independent of the guess itself, which largely
holds for most hashing schemes. 

A pure strategy of the picker and her pure strategy set are the same
as in the Capped-Guesses setting: the picker selects a secret $d$ from
the set of all secrets $\Ps$.
The guesser's strategy, however, can no longer be simply modeled as a
subset of guesses to try, because here, the order of guesses matters:
if the secret is found, the guesser will stop the search and save on
the exploration cost. 
As before, the guesser is strictly better off in expectation to avoid
multiple tries of the same guess. 
Intuitively, a pure strategy of the guesser, as his plan of action,
can be represented as a sequence of guesses without repetition, i.e.,
a \emph{permutation} of a subset of $\Ps$.
The interpretation of a sample strategy $A=\ang{p_1,\ldots,p_{\tau}}$
where $p_i\in\Ps$ for $1\leq i\leq \tau\leq |\Ps|$, will be the
following: first try secret $p_1$ as a guess, if it is not correct,
i.e., if the attempt fails, then try $p_2$, if it fails try $p_3$, and
so on up to $p_{\tau}$, if $p_{\tau}$ fails, then quit the search. 

For any set $\E$, let $\Perm(\E)$ represent the set of all ordered
arrangements (sequences without repetition) of all the members of
$\E$, i.e., $\Perm(\E):=\{A=\ang{a_1,\ldots, a_{|\E|}}|
\{a_1,\ldots,a_{|\E|}\}=\E \}$. 
Moreover, let $\Psi(\E)$ be the set of all permutations over the
elements of the subsets of $\E$, i.e., $\Psi(\E):=\{A|\exists
\E'\subseteq \E \text{ such that } A\in\Perm(\E')\}$.
Using these notations, we can express the strategy space of the
guesser $\A$ as $\A=\Psi(\Ps)$.
Note that the empty sequence, which we denote by $\ang{\texttt{quit}}$
for better presentation, is part of the strategy space of the guesser
as well, representing quitting before making any guesses. 

In \ifcameraready
\else
Appendix~\ref{Appendix:Formal_Model_Derivation_Costly_Guesses}\fi, we
show how the specification of the strategy space of the guesser in
Costly-Guesses scenarios can be formally derived from the standard
game theory models of sequential games with imperfect information.
Specifically, it constitutes the set of \emph{reduced} pure strategies
of a guesser with \emph{perfect recall} (he remembers his past
guesses). 

Given a (pure) strategy profile $(d,A)$, we next compute the utilities
of the two players: $u_D(d,A)$ and $u_A(d,A)$.
First, some notations; we extend the notion of set memberships to
permutations as well, i.e., for a sequence
$A=\ang{a_1,\ldots,a_{\tau}}$, $d\in A$ if and only if
$d\in\{a_1,\ldots, a_{\tau}\}$.
Let $\mathbf{1}_{A}(d)$ be the indicator function determining whether
$d$ appears on sequence $A$, i.e., whether $d\in A$.
Let $\mathrm{pos}_{A}(d)$ refer to the position of the first
appearance of $d$ on sequence $A$ if $d\in A$, and the length of
sequence $A$ otherwise.
For instance, $\mathrm{pos}_{\ang{a,b,c}}(b)=2$ and
$\mathrm{pos}_{\ang{a,b,c}}(e)=3$.
Then we have (compare with \eqref{eq:the_utilities}):
\begin{equation}\label{eq:the_utilities_costly_guesses}
\begin{aligned}
 u_D(d,A)&=-c(d)-\lambda \mathbf{1}_{A}(d)\\
 u_A(d,A)&=\gamma \mathbf{1}_{A}(d)-\sigma \pos_{A}(d)
\end{aligned} 
\end{equation}

As in the Capped-Guesses setting, pure strategies may not be part of
any solution concept, since a pure strategy for the picker translates
to unambiguously revealing her secret.
Hence we should be searching for solutions in the realm of mixed
strategies. 
As before, let $\vec\delta$ and $\vec\alpha$ denote a mix strategy of
the picker and guesser, where $\vec\delta\in\Delta(\Ps)$ and
$\vec{\alpha}\in\Delta(\A)$, with the only difference that $\A$ is now
the set of sequences of distinct guesses, i.e., $\A=\Psi(\Ps)$.
From \eqref{eq:the_utilities_costly_guesses}, the expected utilities
of the players given a mixed strategy profile
$(\vec\delta,\vec\alpha)$ are:

\small
\begin{align*}
&u_D(\vec\delta,\vec\alpha)\!=\!-\sum_{d\in\mathcal{P}} c(d)\vec\delta(d) -\lambda \sum_{A\in\A} \mathbf{1}_{A}(d)\vec\delta(d)\vec\alpha(A)\\
&u_A(\vec\delta,\vec\alpha)\!=\!\gamma\!\sum_{d\in\mathcal{P}}\!\!\sum_{A\in\A}\!\mathbf{1}_{A}(d)\vec\delta(d)\vec\alpha(A) 
\!-\!\sigma\!\!\sum_{d\in\mathcal{P}}\!\!\sum_{A\in\A}\! \pos_{A}(d)\vec\delta(d)\vec\alpha(A) 
\end{align*}
\normalsize
For any $A=\ang{a_i}_i$, we have: $\sum_{d\in\mathcal{P}} \mathbf{1}_{A}(d)\vec\delta(d)=\sum_{i=1}^{|A|}\vec\delta(a_i)$. Moreover: 
$\sum_{d\in\mathcal{P}}\pos_{A}(d)\vec\delta(d)=\sum_{i=1}^{|A|}i\vec\delta(a_i)+|A|(1-\sum_{i=1}^{|A|}\vec\delta(a_i))$. Hence:

\small
\begin{gather}
u_A(\vec\delta,A)\!=\!\gamma\!\sum_{i=1}^{|A|}\vec\delta(a_i)\!-\!\sigma\! \left[\sum_{i=1}^{|A|}i\vec\delta(a_i)\!+\!|A|(1-\sum_{i=1}^{|A|}\vec\delta(a_i))\right] \label{eq:expected_utility_pure_attacker_costly_guesses_simplified}
\end{gather}\normalsize
An alternative method to derive the expression for $u_A(\vec\delta,A)$
is the following: $\sum_{i=1}^{|A|}\vec\delta(a_i)$ is just the
probability that any of the tries on sequence $A$ is the correct
guess. 
Given $\vec\delta$ and $A$, the search reaches $a_i$ in $A$ with
probability $1-\sum_{j=1}^{i-1}\vec\delta(a_j)$. 
Hence, the expected number of tries is
$\sum_{i=1}^{|A|}(1-\sum_{j=1}^{i-1}\vec\delta(a_j))$.
Therefore:
\begin{gather}\label{eq:expected_utility_pure_attacker_costly_guesses_simplified_alternative}
  u_A(\vec\delta,A)=\gamma\sum_{i=1}^{|A|}\vec\delta(a_i)-\sigma
  \sum_{i=1}^{|A|}\left[1-\sum_{j=1}^{i-1}\vec\delta(a_j)\right]
\end{gather}
This is equivalent to the expression in
\eqref{eq:expected_utility_pure_attacker_costly_guesses_simplified}.
In our analysis, we will use either one of the two forms based on
convenience.  

All of the solution concepts introduced in Section
\ref{sec:model:capped_guesses} can be identically defined here as
well. 
We will explore them in detail in the next section. 

\begin{figure}
\begin{framed}
\begin{center}
\textbf{Game 2:} {Costly-Guesses} 
\end{center}
\begin{description}
	\item[Players:] \textsc{\texttt{Picker, Guesser}}
	\item[Strategy Sets:] \textsc{\texttt{Picker's}}: $\{d\in\Ps\}$\\ 
	\textsc{\texttt{Guesser's}}: $\{A|\exists \E\subseteq \Ps \text{ such that } A\in\Perm(\E)\}$
	\item[Utilities:]  \textsc{\texttt{Picker}}: $u_D(d,A)=-c(d)-\lambda \mathbf{1}_{A}(d)$,\\
	\textsc{\texttt{Guesser}}: $u_A(d,A)=\gamma \mathbf{1}_{A}(d)-\sigma \pos_{A}(d)$%
\end{description}
\end{framed}
\label{fig:Game2}
\end{figure}

\section{Analysis of the Costly-Guesses Scenario}\label{sec:Analysis:costly_guesses}
Before we delve into the analysis of the Costly-Guesses scenario, we present a simple yet instrumental lemma: 
\begin{lemma}\label{Lem:Expected_Seeker_Over_Uniform}
 Let $\E$ be a non-empty subset of $\Ps$, and let $\mathrm{unif}(\E)$ represent the uniform distribution over $\E$, i.e., 
 $\vec\delta=\mathrm{unif}(\E)$ if and only if $\vec\delta(p)=\mathbf{1}_{\E}(p)/|\E|$. 
 Then, for any $A\in\Perm(\E)$, $u_A(\mathrm{unif}(\E), A)=\gamma-(|\E|+1)\sigma/2$,  i.e., the expected utility of the guesser for any strategy that exhausts $\E$ is
 $\gamma-(|\E|+1)\sigma/2$.
\end{lemma}
\begin{IEEEproof}
 The secret is a member of $\E$, hence it will be found with certainty, yielding the positive gain of $\gamma$.
 Each guess costs the guesser $\sigma$. The number of guesses before (and including) the correct one is $i$ with probability $1/|\E|$. Hence the expected number of tries is  $\sum_{i=1}^{|\E|}i/|\E|=(|\E|+1)/2$.
\end{IEEEproof}

We will investigate the maximin and minimax strategies of the picker first. 
The picker's maximin strategy is choosing a secret from the cheapest partition, 
i.e., a picking strategy $\vec\delta\in\Delta(\Ps)$ is maximin if and only if 
$\sum_{p\in\E_1}\vec\delta(p)=1$. 
To see this, note that a strategy of the guesser that explores all of the possible secrets, i.e., a permutation of the entire $\Ps$, minimizes the utility of the picker irrespective of the choice of her strategy. Hence, facing this worst case strategy of the guesser, the picker must only select from the cheapest partition.  

A minimax strategy of the picker, on the other hand, is uniform randomization 
over the entire $\Ps$, due to the following two intuitive lemmas:
\begin{lemma}\label{Lem:uniform_the_worst}
 Let $\E$ be a non-empty subset of $\Ps$. Then, for any $\vec\delta\in\Delta(\Ps)$ such that $\supp(\vec\delta)\subseteq\E$, we have: $\sup_{\vec\alpha\in\Delta(\A)}u_A(\vec\delta,\vec\alpha)\geq \sup_{\vec\alpha\in\Delta(\A)}u_A(\mathrm{unif}(\E),\vec\alpha)$.
\end{lemma}
\begin{lemma}\label{Lem:E_1_E_2_uniform}
 Let $\E$, $\E'$ be two non-empty subsets of $\Ps$ such that $|\E|\leq|\E'|$. Then, 
 $\sup_{\vec\alpha\in\Delta(\A)}u_A(\mathrm{unif}(\E),\vec\alpha)\geq \sup_{\vec\alpha\in\Delta(\A)}u_A(\mathrm{unif}(\E'),\vec\alpha)$.
\end{lemma}

The first lemma simply confirms that uniform distribution gives the
least amount of useful information to the guesser.
The second lemma states that uniform randomization over a bigger set
is guaranteed not to help the guesser.
Proof of Lemma~\ref{Lem:uniform_the_worst} is in \ifcameraready
\else
Appendix~\ref{Appendix:Proof_of_Lem:uniform_the_worst}\fi.
Lemma~\ref{Lem:E_1_E_2_uniform} follows directly from
Lemma~\ref{Lem:Expected_Seeker_Over_Uniform}.

As we can see, in the Costly-Guesses setting, the strategically pessimistic and 
the sheer antagonistic plans of action for the picker (her maximin and maximin 
strategies, respectively) lead to uninteresting extremes, suggesting that 
rationality consideration of both players have a more decisive role.  Next, we 
turn our attention to NE solutions.

\subsection{Costly-Guesses: Nash Equilibria}
When $\gamma<\sigma$, the cost of trying even a single guess exceeds the gain of finding the secret. Hence, irrespective of the strategy of the picker, the guesser never enters the game:
\begin{proposition}\label{Prop:Costly_Guesses_NE_very_small_gamma}
 In a Costly-Guesses game, if $\gamma<\sigma$, then in all NE $(\vec{\delta^*},\vec{\alpha^*})$, we have: $\vec{\delta^*}(\E_1)=1$ and 
 $\vec{\alpha^*}(\ang{\texttt{quit}})=1$, i.e., the picker chooses from the cheapest partition and the guesser does not make any attempt.
\end{proposition}

What happens when $\gamma>\sigma$?
If $\gamma<(1+\sum_{i=1}^M |\E_i|)\sigma/2$ for some $M\leq N$, then
following Lemma~\ref{Lem:Expected_Seeker_Over_Uniform}, the picker can
dissuade the guesser from entering the game by uniformly randomizing
over the first $M$ partitions.
When the picker assumes a high cost for losing her secret, i.e., for
large values of $\lambda$, this seems to be something she will opt
for.
However, our next proposition reveals that, surprisingly, if there is
no partition that is big enough that uniform randomization over it
\emph{alone}, i.e., single-handedly, can dissuade the guesser from
entering, then in \emph{all} NE of the game, the picker chooses a
cheapest secret and loses it with certainty, and remarkably, this is
true \emph{irrespective} of the magnitude of $\lambda$:
\begin{proposition}\label{Prop:Costly_Guesses_NE_big_gamma}
  In a Costly-Guesses game, if $\gamma>(1+|\E_i|)\sigma/2$ for all $i$
  for which $C_i<C_1+\lambda$, then in all NE
  $(\vec{\delta^*},\vec{\alpha^*})$, we have: $\vec{\delta^*}(\E_1)=1$
  and $u_D(\vec{\delta^*},\vec{\alpha^*})=-C_1-\lambda$, i.e., the
  picker chooses only from the cheapest partition and the guesser
  finds it with certainty.
\end{proposition}

The detailed proof of the proposition is provided in \ifcameraready
\else
Appendix~\ref{Appendix:Proof_of_Prop:Costly_Guesses_NE_big_gamma}\fi.
Here we provide an informal summary of the proof with the aim of
giving an idea why we have this ``failure'' of NE for the picker: in
any NE, the mixed strategies of the two players must be best responses
to each other.
Therefore, in a NE, the picker only assigns positive probability of
selection from costlier partitions because of the threat imposed by
the exploration probabilities of the guesser.
Suppose there is a NE in which the picker assigns strictly positive
probabilities to secrets from partitions $\E_1$ to $\E_M$.
This means that the guesser explores $\E_1$ to $\E_M$ with strictly
decreasing probabilities.
This in turn implies that the guesser must find it a best response to
explore $\E_{M-1}$ and not $\E_{M}$ among his set of best responses
that he randomizes over.
Note that the picker never assigns a strictly higher probability to
members from a costlier partition.
This means that if exploring $\E_{M-1}$ and not $\E_{M}$ must be a
best response of the guesser, so must be exploring $\E_1$ through
$\E_{M-1}$ and not $\E_M$.
However, this can never be the case: if the guesser explores all of
the partitions $\E_1$ to $\E_{M-1}$ and fails, then given the
randomization of the picker, he is now certain that the secret is in
$\E_M$.
Given the condition $\gamma>(\E_M+1)\sigma/2$, the guesser is strictly
better off to continue to explore $\E_M$ as well.
Hence, the starting assumption about the NE strategy of the picker
could not be true.  

The next proposition shows what may happen when the condition of
Prop.~\ref{Prop:Costly_Guesses_NE_big_gamma} is relaxed (proof
is in \ifcameraready\else
Appendix~\ref{Appendix:Proof_of_Prop:Costly_Guesses_NE_average_gamma}\fi):
\begin{proposition}\label{Prop:Costly_Guesses_NE_average_gamma}
  In a Costly-Guesses game where $\gamma>\sigma$, if $\exists
  M=\min\{i|\gamma<(|\E_i|+1)\sigma/2,\ C_i\leq C_1+\lambda\}$,
  then 
  in all NE $(\vec{\delta^*}\!,\vec{\alpha^*})$ we have:
  $u_D(\vec{\delta^*},\vec{\alpha^*})\leq -C_M$.
\end{proposition}

This proposition does not quite redeem the stark situation with NE
solutions for the picker.
For instance, consider a case where the picker could prevent the
guesser from entering the game by randomizing over $\E_1$ and $\E_2$,
and the cheapest partition that is big enough to single-handedly
prevent the guesser from entering the game is $\E_3$. 
Then the picker has to settle for a cost of $C_3$, which can be much
larger than any weighted average of $C_1$ and $C_2$.
Moreover, the proposition only provides a (tight) upper-bound on the
expected utility of the picker among all NE.
That is, $-C_M$ is the expected utility of the picker in the best NE
for her, and worse NE for the picker can still exist.
In particular, if $\gamma>(|\E_1|+1)\sigma/2$, then
$(\vec{\delta^*},\vec{\alpha^*})$ where
$\vec{\vec{\delta^*}}=\unif(\E_1)$ and
$\vec{\alpha^*}=\unif\left(\Perm(\E_1),\Perm(\cup_{i=2}^N\E_i)\right)$
is also technically a NE: given that the picker chooses uniformly from
the cheapest partition, it is a best response for the guesser to
explore the whole set of secrets starting from the cheapest partition;
likewise if the guesser's strategy is to explore the whole set of
secrets, then the guesser's best response is to choose from the
cheapest partition, since she will lose her secret anyway.
This NE, as in Prop.~\ref{Prop:Costly_Guesses_NE_big_gamma},
yields for the picker the worst possible in any NE: her maximin
utility, that is $-C_1-\lambda$.

What causes the poor performance of the picker in NE is the absence of a credible commitment to a deterring randomization. 
Indeed the picker prefers to induce the guesser to abstain, however, if the guesser is not going to enter the game, the picker prefers to select a least costly secret. The picker can remove this possibility from the reasoning of the guesser by credibly communicating a commitment to a mixed strategy. This is exactly the setup for Strong Stackelberg Equilibria, which we analyze next. 

\subsection{Equilibrium Example: Passwords}\label{sec:example-costly-password-analysis}
\begin{figure}
\includegraphics[width=0.95\linewidth]{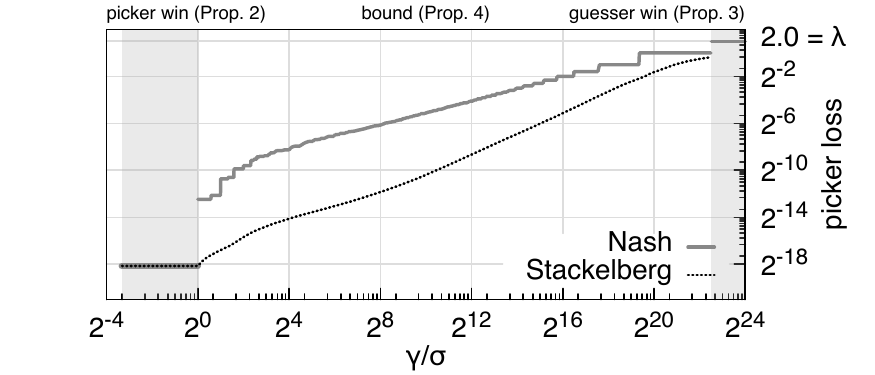}
\caption{\label{fig:pass-costly-nash-stack} RockYou-based password
  picker loss in Nash equilibrium or lower bound (gray) and in
  Stackelberg equilibrium (black dotted) as function of
  $\gamma/\sigma$.
  For all, $ \lambda = 2 $.}
\end{figure}

Figure~\ref{fig:pass-costly-nash-stack} shows the result of equilibrium behavior
on the loss of the picker in the costly guesses model for password
selection.
Most of the figure is a lower bound on loss as per
Prop~\ref{Prop:Costly_Guesses_NE_very_small_gamma}.
For low ratios of $ \gamma/\sigma $, the guesser does not participate
at all and results only in the cost of picking the simplest password.
For large enough ratios, the picker gives up, incurring a loss of $
\lambda $ and the cost of the simplest password.
In the mid-range cases, the loss factor $ \lambda $ plays no role. 

\subsection{Equilibrium Example: Cryptographic
  Keys}\label{sec:example-costly-keys-analysis}

\begin{figure}
  \includegraphics[width=0.95\linewidth]{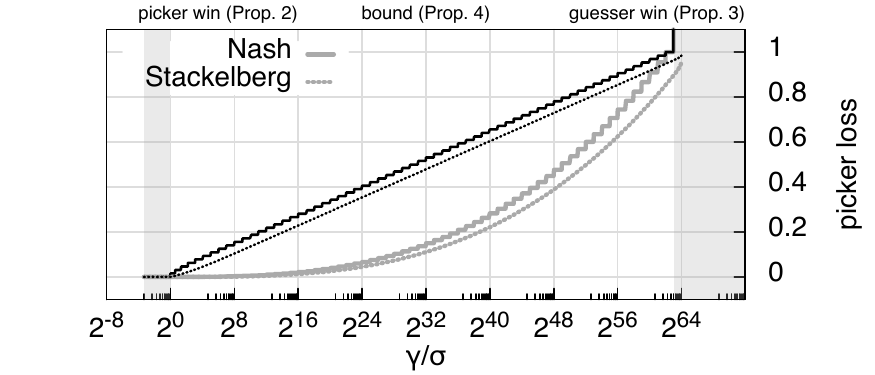}
  \caption{\label{fig:keys-costly-nash-stack} Picker loss in the Nash
    (solid lines) and Stackelberg (dotted lines) key picking strategy
    as a function of $ \gamma/\sigma $, for linear and cubic cost
    functions (lighter line is cubic cost).
    For all, $ \lambda = 2 $.}
\end{figure}


The solid lines in Figure~\ref{fig:keys-costly-nash-stack} show picker
loss for the key selection scenario with costly guesses.
The results contain essentially the same features as the password
selection example: low-enough ratios of $\gamma/\sigma$ results in
guesser not participating, high enough ratios result in the picker
giving up (this is out of the frame at picker loss equal to $\lambda =
2 $), and in between the picker exhibits an increasing expected loss.

\subsection{Costly-Guesses: Strong Stackelberg Equilibria}
Here we assume the picker has access to an apparatus that enables her
to credibly communicate a commitment to a mixed strategy to the
guesser. 
We develop the optimal randomizations for the picker given the fact
that the guesser, observing the committed randomization, best-reacts
to it.
Formally, we derive the SSE strategies of the picker. 

First, note that if $\gamma<\sigma$, then irrespective of the choice
of the picker, the guesser will never attempt a guess.
Then the SSE strategy of the picker for these cases is, trivially, a
choice from the cheapest partition, yielding the picker a utility of
$-C_1$ and the guesser, zero.
Therefore, in the rest of this section, we only consider
$\gamma>\sigma$.
We show the following: if at all worth protecting the secret, the
picker should commit to a randomization that makes not entering the
game a best response for the guesser, i.e., the cheapest randomization
that leaves the guesser indifferent between entering the game and
quitting at the beginning.
In particular, committing to randomizations that leave incentive for
the guesser to perform even a partial search is never
optimal.\footnote{This is reminiscent of this pithy quote
  \cite{mcadams2014game} from Zhuge Liang, a recognized ancient
  Chinese military strategist and statesman: ``\textit{The wise win
    before they fight, while the ignorant fight to win.}''} 
We specifically develop a linear optimization that gives the SSE
strategy of the picker.
%
\begin{proposition}\label{Prop:Costly_Guesses_Stackelberg}
 Consider the following linear programming:
 \small
 \begin{align*}
 &u^{*}_{D}=\max_{\nu_i} \left[-\sum_{i=1}^N C_i \nu_i \right]\ \ \mathrm{subject\ to:} \\
 &\nu_i\!\geq\! 0\ \mathrm{for}\ 1\!\leq\! i\!\leq\! N,\ \sum_{i=1}^N \nu_i=1,\ \frac{\nu_i}{|\E_i|}\!\geq\! \frac{\nu_{i+1}}{|\E_{i+1}|}\ \mathrm{for}\ 1\!\leq\! i\!\leq\! N-1\\
 & \gamma\sum_{i=1}^K \nu_i \!-\! \sigma\sum_{i=1}^K\left[|\E_i|(1-\sum_{j=1}^{i-1}\nu_j)-\frac{|\E_i|-1}{2}\nu_i\right]\!\leq\! 0\ \mathrm{for}\ 1\!\leq\! K\!\leq\! N
 \end{align*}\normalsize
For $(|\Ps|+1)\sigma/2>\gamma$, the LP is feasible. Let 
$(\nu^*_1,\ldots,\nu^*_N)$ be a solution.  If $u^*_D>-C_1-\lambda$, then a SSE 
strategy of the picker is $\delta(p)=\nu^*_i/|\E_i|$ for $p\in\E_i$. 
If $u^*_D<-C_1-\lambda$, the SSE strategy of the picker is to simply choose a 
secret from the cheapest partition (which induces the guesser to enter, explore 
that partition and find the secret with certainty). Same is true when 
$(|\Ps|+1)\sigma/2<\gamma$.\footnote{One can find uniqueness conditions for the 
SSE strategy of the picker, using standard results in linear programming (e.g. 
\cite{mangasarian1979uniqueness}). However, the uniqueness of the utility of the 
picker follows from the optimization itself.} 
\end{proposition}

The proof of the proposition is provided in \ifcameraready
\else 
Appendix~\ref{Appendix:Proof_of_Prop:Costly_Guesses_Stackelberg}.\fi  
Note that when $(|\Ps|+1)\sigma/2<\gamma$, following
Lemma~\ref{Lem:Expected_Seeker_Over_Uniform}, even uniform
randomization over the entire set of $\Ps$ does not deter the guesser
from entering the game and exploring the whole secret space, as it
yields him a strictly positive utility of $\gamma-(|\Ps|+1)\sigma/2$.
Since uniform randomization is a minimax strategy of the picker
(intuitively, it gives the least useful information to the guesser),
any other randomization also results in a strictly positive utility
for full exploration of the guesser.
This means the best strategy of the picker is then choosing from a
cheapest partition, since she will lose her secret to the guesser
anyway. 

When $(|\Ps|+1)\sigma/2>\gamma$, uniform randomization over a subset
of secrets can lead to a negative expected utility of the guesser for
entering the game and exploring any portion of the secret space.
However, our numerical examples of the proposition reveal that the
cheapest randomization that achieves this goal is almost never
completely uniform (or even necessarily uniform over the union of some
cheapest partitions except for the costliest of them). 

%

\subsection{Stackelberg Examples}\label{sec:example-costly-keys-stack}

\begin{figure}
\centering
  \includegraphics[width=0.95\linewidth]{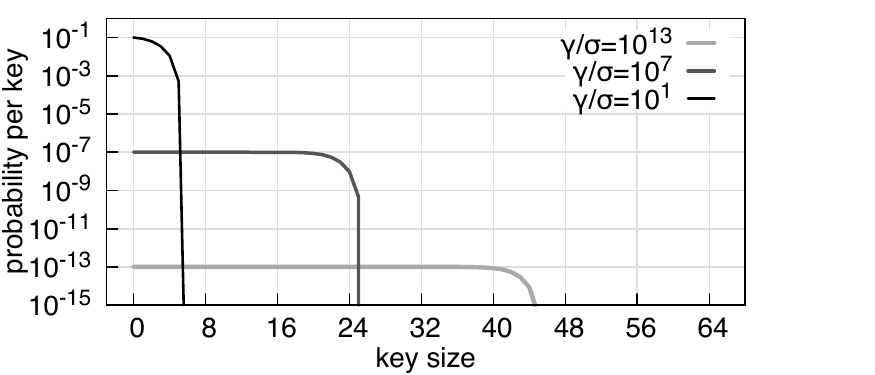}
  \caption{\label{fig:keys-stack} The distribution of key sizes in the
    Stackelberg key picking strategy for a variety of $\gamma/\sigma$
    values.}
\end{figure}

The difference between the Nash equilibrium and the Stackelberg
equilibrium is demonstrated in Figure~\ref{fig:pass-costly-nash-stack}
for the password picking example and in
Figure~\ref{fig:keys-costly-nash-stack} for the key selection example.
In both, the picker's loss in Stackelberg equilibrium as a function of
$\gamma/\sigma$ is denoted by dotted lines.
In the case of key selection, linear and cubic cost models are shown
with linear as dark lines and cubic as light lines.
The Stackelberg strategies can be seen to perform better than the Nash
strategies shown as solid lines.

Figure~\ref{fig:keys-stack} demonstrates the Stackelberg strategy for
key selection in more detail for three different $\gamma/\sigma$
values.
The selection of key in these solutions is mostly uniform among keys
up to a certain length except for larger keys whose probability
sharply falls off.
The range of key sizes which are selected over increases with
$\gamma/\sigma$.


\section{Discussions}\label{sec:Discussions}
\textbf{Policy implications of picker's optimal strategies in Capped-Guesses:} 
System administrators usually use password selection rules (composition policies) to increase the entropy of the passwords selected by users. 
The choice of an optimal rule-set has been a topic of research \cite{komanduri2011passwords,kelley2012guess,bonneau2012science,weir2010testing}. 
Recall that an interpretation of the partitions of the secret space based on their usability costs was that these partitions can be assumed to satisfy increasingly complex password composition rules. 
Hence, our Prop.~\ref{Prop:NE} suggests that the search for the ``optimal'' 
composition rule is amiss. The optimal secret picking strategy is a uniform 
randomization over a ``union'' of these partitions, and hence, no single 
composition rule is optimal. Our result suggests that optimal composition rules 
are generated by randomizing across different rule-set, and specifically, each 
composition rule should be prompted to a user with a probability that is 
proportional to the size of password space created by that rule.

\hide{In \cite{kelley2012guess}, e.g., the authors empirically measure the
strength (guessability) of passwords created by a number of different composition rules typically used against some of the state-of-the-art cracking algorithms.
In particular, they empirically found that `` a password-composition
policy requiring long passwords with no other restrictions
provides (relative to other policies we tested) excellent
resistance to guessing.''
Our analysis in part sheds light on why this rather counter-intuitive phenomenon 
is the case: imposing no restrictions other than a long password is in fact the 
closest to  a uniform randomization. 
We expect that a yet better policy is a randomization between a group of 
policies. 
}

\textbf{Credible commitment to a randomization}
Recall that in a Capped-Guesses scenario, the SSE, NE and Maximin strategies of the picker turn out to be identical. In particular, there is no gain in communicating a commitment to the adversaries. 
In the Costly-Guesses model, however, a credible commitment to a randomization makes a substantial difference, and is critical to prevent the failure of NE. 
Hence, in such situations, it is not sufficient to have access to a randomization device, but further, the randomization should be made public knowledge and verifiable to become credible. 

\textbf{Optimal attacks in Capped-Guesses} Our Prop.~\ref{Prop:NE} suggest 
that in a Capped-Guesses attack, e.g. using pre-computed tables, even facing a
rational defender that plays optimally and hence uses uniform randomization, 
the adversary must choose passwords randomly from the whole selection range of 
the user, however, should choose simpler passwords with more probability and 
include more difficult ones with increasingly less probabilities. 

\textbf{Interpretation of mixed strategies} The game theoretic solutions that we developed involved randomization. Specifically, in mixed NE, each player's randomization leaves the other indifferent across his/her randomization support. Although these behaviors can be explicitly associated with deliberate randomization or through the use of randomization devices (e.g. when a random key generator algorithm is used), these are not the only way such equilibria can be interpreted. Without going to the details \cite{herbert2000game}, we just mention some of the alternative interpretations equilibrium solution involving mixed strategies. 
Namely, the probabilities can represent (a)~time averages of player's behavior that exploit an ``adaptive'' process, (b)~fractions of the total ``population'' of each player that adopt pure strategies, (c)~limits of pure strategy Bayesian equilibria where each player is slightly uncertain about the payoffs of the others, and
(d)~A ``consistent'' set of ``beliefs'' that each player has about the other regarding their behavior. 

\textbf{Other applications:} Finally, it is worth mentioning that even though we 
motivated our models based on password and cryptographic key selection, 
the generality of the model allows it to be applicable to other contexts as 
well. As an example of a completely different context but with identical 
abstraction, consider a user that aims to send a convoy from a source to a 
destination over a transport network, or transmit a packet over a communication 
network. There are multiple paths available and the user's objective is to use 
this path diversity to minimize the risk of being intercepted on a path by an 
adversary.
However, the paths may have different utilities as some may  provide lower delays and higher quality of service, a preference that can be exploited by adversaries as well.

\section{Related Work}\label{sec:Related_Work}

User password selection and attacks has been extensively studied in the literature
\cite{komanduri2011passwords,bonneau2012guessing, bonneau2012science, weir2010testing,
  kelley2012guess,ur2012does}, and due to its practical significance, continues 
to be a hot area of research \cite{florencio2014administrator}.
These works generally aim at evaluating the efficacy of password attacks as 
well as measuring the strength of different password composition rules through  
statistical metrics. In contrast to our work, these papers consider the user or 
the adversaries one at a time, 
as opposed to considering that both parties 
will adapts to each other's choice of policies. Analysis of such strategic 
actions and reactions can be
done through a game theoretic framework, which to our best of knowledge, our 
work is the first in this context. 

Game and decision theory has been applied in other cybersecurity 
contexts with promising potentials \cite{tambe2011security,alpcan2010network}. 
The first part of our work (Capped-Guesses) is, in its abstract form, similar to 
the security game model analyzed in \cite{yin2010stackelberg}. 
In their model, the defender has limited resources to cover a wide range of 
targets, while an adversary chooses a single target to attack. If targets are 
thought of as secrets, the defender in their model is akin to the guesser in our 
work, and their adversary is our picker. 
Therefore, our Capped-Guesses model is the ``complement'' of their model. 
Specifically, the results that they develop 
for their defender will be translatable to our guesser. However, the focus of 
our paper was on the picker. 

Another line of research from theoretical game theory is search theory and 
search games \cite{alpern2013search}. 
Existence of user preferences over the secrets to pick from is missing from such 
models. However, such preferences are at the heart of usability-security 
trade-off settings investigated in our paper. 


\section{Conclusion}\label{sec:Conclusion}
We developed tractable game-theoretic models that capture the essence of secret picking vs guessing attacks in the presence of preferences over the secret space. 
We then provided a full analysis of our models with the aim of investigating fundamental trends and properties in the design of secret-picking policies that attain optimal trade-offs between usability and security, taking into account the exploitation of the knowledge of such trade-offs by an adversary.  
Notably, we computed the secret picking policies that are optimal with respect to a range of strategic metrics (Maximin, Minimax, Nash Equilibria, Stackelberg Equilibria).

We distinguished between two classes of guessing attacks: those in which the number of available guesses to an adversary is capped (Capped-Guesses), and those in which an adversary has potentially unlimited number of tries but incurs a cost per each guess (Costly-Guesses).  Our analysis revealed the crucial role that such distinction between the nature of the guessing adversary plays on the expected outcome. 
Specifically, we showed that in the Capped-Guesses settings, the NE strategy of the secret picker is still uniform but over a low-cost subset of the secret space, where the size of the subset depends on the parameters of the adversary only through the number of available guesses. 
In contrast, we established that for Costly-Guesses scenarios, except for trivial cases, NE fails to attain a desirable outcome for the secret picker. For this setting, we showed how deterrence of adversaries as her optimal strategy crucially depend on existence of a credible commitment to a randomization strategy. We illustrated our results through a series of numerical examples using real-world data-sets.

\paragraph*{Future Directions}
One of the main areas of extending this work is dealing with uncertainty in the 
parameters of the players. For instance, the picker may not accurately know the 
type of the guesser or their guessing size cap or their guessing costs. 
One approach to formally take such uncertainties into account is a Bayesian 
game approach, for which, this work lays the foundation of. 

Moreover, in this paper, we assumed that once the secret is selected, the picker
does not get to change it later, either as a blind (open-loop) policy
or as a reaction to some signal generated by the actions of an
adversary.
Note that if the act of changing the secret does not bring any cost to
the picker, and both parties are aware of secret-changing occasions,
then our results are still applicable, since in essence, the two
players play the same game after each reset. 
However, the previous choices of the picker may affect her future
utilities, and hence the whole game.
For instance, the act of changing the secret may be costly for the
picker, or changing the secret only slightly may be associated with
less cost than changing it drastically.
In such scenarios, a rational adversary can exploit such preferences
and carry some useful information from each round of the game to boost
his overall attack.
Investigation of such scenarios using dynamic game theory is a
potential extension of our work.


Another interesting scenario to investigate is when the picker is choosing multiple secrets, where there is a increasing loss for the number of secrets guessed correctly by an adversary. 
The two extremes are (1)~when the guesser wins if any of the secrets are discovered, and (2)~when the guesser wins only if all of the secrets are discovered. 

\section{Acknowledgments}
This research was partially sponsored by US Army Research laboratory and the UK 
Ministry of Defence under Agreement Number W911NF-06-3-0001. The views and 
conclusions contained in this document are those of the authors and should not 
be interpreted as representing the official policies, either expressed or 
implied, of the US Army Research Laboratory, the U.S. Government, the UK 
Ministry of Defence, or the UK Government. The US and UK Governments are 
authorized to reproduce and distribute reprints for Government purposes 
notwithstanding any copyright notation hereon.
The first and third authors were also supported by the Project ``Games and 
Abstraction: The Science of Cyber Security'' funded by EPSRC, Grants: 
EP/K005820/1, EP/K006010/1.

\bibliographystyle{IEEEtran}

\ifcameraready
\else
\appendices
\section{Proof of Proposition~\ref{Prop:NE}: NE in Capped-Guesses}\label{Appendix:Proof_of_Prop:NE}
\begin{IEEEproof}
First, consider the ``ordinary'' cases. We provide the proof in two steps:
  
  \textbf{First step:} We show that the proposed mixed strategies do indeed correspond to legitimate probability distributions over pure strategy spaces.
  For $\mathbf{\vec\delta}^*=\unif(\cup_{i=1}^J\E_i)$, this is trivially true.
  For $\vec{\alpha^*}$, recalling $\sum_{A\in\A}\vec{\alpha^*}(A)\mathbf{1}_{A}(p)=\vec{\rho^*}(p)$, consistency of the conditions in \eqref{eq:CaseN_alpha*} with legitimate probability distributions translates to establishing two facts: (a)~$\vec{\rho^*}(p)\geq0$ for all $p\in\Ps$, and (b)~$\sum_{p\in\Ps}\vec{\rho^*}(p)\leq K$, as we do next.

   (a)~Showing $\vec{\rho^*}(p)\geq0$ for all $p\in\Ps$: For $p\in \E_i$ where $i>J$, $\vec{\rho^*}(p)=0$. For $p\in \E_i$ where $i\leq J$, suppose there exists a $p\in\E_l$ where $1\leq l \leq J$, such that the expression in \eqref{eq:Alpha^*_below_J} is strictly negative:
   \begin{multline*}
    \vec{\rho^*}(p)=\frac{K}{\sum_{j=1}^{J}|\E_j|}+\frac{\sum_{j=1}^{J}C_j|\E_j|-C_l\sum_{j=1}^{J}|\E_j|}{\lambda \sum_{j=1}^J|\E_j|}\\
    =\frac{\lambda K +\sum_{j=1}^J C_j|\E_j|-C_l\sum_{j=1}^J|\E_j|}{\lambda \sum_{j=1}^{J}|\E_j|}<0
   \end{multline*}
   Since $C_l\leq C_J$, this in turn implies:
${\lambda K+\sum_{j=1}^J C_j|\E_j|}<C_J{\sum_{j=1}^J|\E_j|}$, 
which is in contradiction with the specification of $J$ as $\max\mathcal{J}$, where $\mathcal{J}$ is defined in \eqref{eq:J_Definition}. The claim hence follows. 

   (b)~Showing $\sum_{p\in\Ps}\vec{\rho^*}(p)\leq K$: Directly from \eqref{eq:CaseN_alpha*}, we have:
  \[
    \sum_{p\in\Ps}\vec{\rho^*}(p)
    =    K\sum_{i=1}^J\frac{|\E_i|}{\sum_{j=1}^J|\E_j|}+\sum_{i=1}^J|\E_i| B_i=K+\sum_{i=1}^J |\E_i| B_i
   \]
   In the following, we show that $\sum_{i=1}^JB_i=0$:
   \begin{multline*}
    \sum_{i=1}^J|\E_i| B_i= \sum_{i=1}^J|\E_i|\dfrac{\sum_{j=1}^{J}C_j|\E_j|-C_i\sum_{j=1}^{J}|\E_j|}{\lambda  \sum_{j=1}^J|\E_j|}\\=
    \frac{\sum_{i=1}^J|\E_i|\sum_{j=1}^{J}C_j|\E_j|-\sum_{i=1}^J|\E_i|C_i\sum_{j=1}^{J}|\E_j|}{\lambda  \sum_{j=1}^J|\E_j|}=0
   \end{multline*}
   
The above in fact enables an interesting observation: Since $\sum_{i=1}^J|\E_i|B_i=0$, the value of  $B_i$ must be positive for some instances of $i$, and must be negative for others. 
From \eqref{eq:CaseN_alpha*}, recall that $B_i$ represents the ``bias'' of the guesser in exploring the $i$'th partition, away from the uniform selection of the picker. Hence, the guesser explores the partition $\E_i$ that has positive (resp., negative) $B_i$ with higher (resp., lower) probability than uniform random selection.
Referring to the expression of $B_i$, we have $\sgn(B_i)=\sgn\Big[\big({\sum_{j=1}^JC_j|\E_j|}/{\sum_{j=1}^J|\E_j|}\big)-C_i\Big]$. Hence, $B_i$ is positive (resp., negative), if among the partitions that the picker randomizes over, the cost of choosing a secret from $\E_i$ is less than (resp., more than) the overall average cost of choosing a secret. In words, in the NE described by the proposition, \emph{the guesser explores the relatively favored partitions of the picker with a positive bias compared to the relatively less favored partitions}. Note, particularly, that this biased exploration is \emph{not} because the guesser believes that there is a higher chance of having a correct guess from those partitions: he searches them more \emph{despite} knowing that the secret is equally likely to be from any of the partitions $\E_1$ to $\E_J$. The rationale of this strategy becomes clear in the next step of the proof.

\textbf{Second Step:} The ``Nash'' property: given the strategy of the picker, the guesser plays his best response, and vice versa.

(a)~That the guesser's strategy is a best response to the picker's is simple to establish: When the guesser adopts $\vec{\delta^*}=\unif(\cup_{i=1}^J\E_i)$, the secret is not from the set $\cup_{i=J+1}^N\E_i$. Hence, given the fact that for ``ordinary'' cases $K<\sum_{i=1}^J|\E_i|$, the guesser should not ``waste'' any of his guesses by choosing from $\cup_{i=J+1}^N\E_i$, hence the property in \eqref{eq:Alpha^*_beyond_J}. Moreover, given $\vec{\delta^*}$, the secret is equally likely to be any of the members of the set $\cup_{i=1}^J\E_i$. Hence any distribution of choosing $K$ guesses over $\cup_{i=1}^J\E_i$ is a best response, including the ones that satisfy the proposed property in \eqref{eq:Alpha^*_below_J}. 

(b)~Similarly, we show that the picker's strategy is a best response to that of the guesser's. First of all, note that the secrets from the same partition has the same probability of being on the guess dictionary of the guesser, and they also all share the same choosing cost. Hence, any redistribution of the picker's probability within the same partition is a best response, including a uniform one. Next, we show that the guesser's strategy makes the picker indifferent in choosing of her secret from any of the sets $\E_1$ to $\E_J$: 
  Indeed, the description of $\vec{\rho^*}$ in \eqref{eq:Alpha^*_below_J} is by construction such that the utility of the picker for choosing her secret from set $\E_i$, for any $i\leq J$, is equal to $-\big(\sum_{j=1}^JC_j|\E_j| +\lambda K\big)/\big(\sum_{j=1}^J |\E_j|\big)$ (which is in fact equal to $u_D(\vec{\delta^*},\vec{\alpha^*})$).
To see this, first note that, given $\vec{\alpha^*}$, the utility of the picker for choosing secret $p\in\E_i$ for $i\in\{1,\ldots,J\}$ is: $-C_i-\lambda\vec{\rho^*}(p)$.  Now, enforcing: 
  \begin{gather*}
   -C_i-\lambda\vec{\rho^*}(p)= -\frac{\sum_{j=1}^JC_j|\E_j| +\lambda K}{\sum_{j=1}^J |\E_j|}
  \end{gather*}
after simplification, yields exactly the description of $\vec{\rho^*}$ in \eqref{eq:Alpha^*_below_J}. 
Now consider any deviation from $\vec{\delta^*}=\unif(\cup_{i=1}^J\E_i)$ and call it $\vec\delta'$.
We have:
\begin{multline}
 u_D(\vec\delta',\vec{\alpha^*})= -\frac{\sum_{j=1}^JC_j|\E_j| +\lambda K}{\sum_{j=1}^J |\E_j|}\vec\delta'(\cup_{i=1}^J\E_i)\\
 -\sum_{i=J+1}^NC_i\vec\delta'(\E_i)
 \label{eq:utility_deviation_delta}
\end{multline}
On the other hand, we have:
$ u_D(\vec{\delta^*},\vec{\alpha^*})= -\big(\sum_{j=1}^JC_j|\E_j| +\lambda K\big)/\big(\sum_{j=1}^J |\E_j|\big)$.
Following the definition of $J$ as $\max \mathcal{J}$, where $\mathcal{J}$ is given in \eqref{eq:J_Definition}, we have: $-\big({\sum_{j=1}^JC_j|\E_j| +\lambda K}\big)/\big({\sum_{j=1}^J |\E_j|}\big)>-C_i$ for $i=J+1$, and hence, for all $i\geq J+1$ (since $C_i$s are strictly increasing). Referring to \eqref{eq:utility_deviation_delta}, this establishes that $ u_D(\vec\delta',\vec{\alpha^*})\leq  u_D(\vec{\delta^*},\vec{\alpha^*})$, which is the Nash property for the picker.

Now we turn our attention to the ``total defeat'' cases. Note that $\vec{\rho^*}(p)=1$ for all $p\in\E_1$. 
Hence, given the strategy of the picker, the guesser's strategy is a best response. 
On the other hand, given the strategy of the guesser, the picker does not have an incentive to deviate from her strategy either: Consider a deviation $\vec\delta'$. Then $u_D(\vec\delta',\vec{\alpha^*})\leq -(C_1+\lambda)=u_D(\vec{\delta^*},\vec{\alpha^*})$, hence, there is no benefit for the picker to unilaterally deviate either. All that remains to show is the feasibility of the guesser's strategy. Note that: $\sum_{d\in \cup_{i=1}^J \E_i}[1-(C_i-C_1)/\lambda]= \sum_{i=1}^J |\E_i| + C_1\sum_{i=1}^J|\E_i|/\lambda - \sum_{i=1}^J 	C_i|\E_i|/\lambda <K$. The latter follows from the fact that for ``total defeat'' cases, 
we have: $C_1+\lambda < (\lambda K + \sum_{i=1}^J C_i|\E_i|)/(\sum_{i=1}^J |\E_i|)$.
\end{IEEEproof}

\section{Proof of Lemma~\ref{Lem:NE_is_Maximin}: Picker's NE-Maximin Equivalence in Capped-Guesses}\label{Appendix:Proof_of_Lem:NE_is_Maximin}
\begin{IEEEproof}
The proof is immediate once we note from \eqref{eq:exptec_utilities_mixed} that for any $\vec\delta\in\Delta(\Ps)$, $\vec{\alpha^*}\in\arg\max_{\vec\alpha\in\Delta(\A)} u_{A}(\vec\delta,\vec\alpha)$ if and only if:  
$\vec{\alpha^*}\in\arg\min_{\vec\alpha\in\Delta(\A)} u_{D}(\vec\delta,\vec\alpha)$.
This is because: $u_{D}(\vec\delta,\vec\alpha)=-(\lambda/\gamma) u_{A}(\vec\delta,\vec\alpha)+\phi(\vec\delta)$, where $-(\lambda/\gamma)<0$ and $\phi(\vec\delta)$ is an expression that does not depend on $\vec\alpha$.\\
For any $\vec\delta$, recall the notations of $\underline{u_D}(\vec\delta):=\min_{\vec\alpha\in\Delta(\A)}u_{D}(\vec\delta,\vec\alpha)$, and $\underline{u_D}^{\mathrm{max}}:=\max_{\vec\delta\in\Delta(\Ps)}\underline{u_D}(\vec\delta)$. Then, following the definition, $\vec{\delta^*} \in\Omega^{\mathrm{D}}_{\mathrm{Maximin}}$ if and only if $\underline{u_D}(\vec{\delta^*})=\underline{u_D}^{\mathrm{max}}$. 
Suppose $(\vec{\delta^*},\vec{\alpha^*})$ is a NE.
Following the discussion above, in the Capped-Guesses game we have: $u_D(\vec{\delta^*},\vec{\alpha^*})=\underline{u_D}(\vec{\delta^*})$. 
By definition, (A):~$\underline{u_D}^{\mathrm{max}}=\max_{\vec\delta\in\Delta(\Ps)}\underline{u_D}(\vec\delta)\geq \underline{u_D}(\vec{\delta^*})$.
On the other hand, since $\vec{\delta^*}$ is a best response to $\vec{\alpha^*}$, as required in a NE, we have:
$u_{D}(\vec{\delta^*},\vec{\alpha^*})\geq u_{D}(\vec\delta,\vec{\alpha^*})\geq \underline{u_{D}}(\vec\delta)$ for any $\vec\delta\in\Delta(\Ps)$, where the latter inequality follows from  the definition of $\underline{u_{D}}(\vec\delta)$. 
Hence, (B):~ $u_{D}(\vec{\delta^*},\vec{\alpha^*})\geq \max_{\vec\delta\in\Delta(\Ps)} \underline{u_{D}}(\vec\delta)=\underline{u_D}^{\mathrm{max}}$. 
Putting inequalities (A) and (B) together, we obtain $u_{D}(\vec{\delta^*},\vec{\alpha^*})=\underline{u_D}^{\mathrm{max}}$, that is, $\vec{\delta^*}\in\Omega^{\mathrm{D}}_{\mathrm{Maximin}}$.

Now, consider the other direction: Suppose $\vec\delta'\in\Omega^{\mathrm{D}}_{\mathrm{Maximin}}$. Then there exists an $\vec\alpha'\in\Delta(\A)$ such that
$u_D(\vec\delta',\vec\alpha')=\underline{u_D}^{\mathrm{max}}$ and $u_D(\vec\delta',\vec\alpha')=\underline{u_D}(\vec\delta')$. 
The latter in our game implies that $u_A(\vec\delta',\vec\alpha')\geq u_A(\vec\delta',\vec\alpha)$ for any $\vec\alpha\in\Delta(\A)$, hence $\vec\alpha'$ is a best response to $\vec\delta'$.
Also, the former implies that $u_D(\vec\delta',\vec\alpha')\geq \underline{u_D}(\vec\delta)\geq u_{D}(\vec\delta,\vec\alpha')$ for any $\vec\delta\in\Delta(\A)$, hence $\vec\delta'$ is a best response to $\vec\alpha'$ too. Therefore, $(\vec\delta',\vec\alpha')$ is a NE and $\vec\delta'\in \Omega^{\mathrm{D}}_{\mathrm{NE}}$. 
\end{IEEEproof}

\section{Proof of Corollary~\ref{Corr:Interchangeability_I}: Capped-Guesses NE Interchangeability I}\label{Appendix:Proof_of_Corr:Interchangeability_I}
\begin{IEEEproof}
 Since $(\vec{\delta^*}_1,\vec{\alpha^*}_1)$ is a NE, we have: $u_D(\vec{\delta^*}_1,\vec{\alpha^*}_1)\geq u_D(\vec\delta',\vec{\alpha^*}_1)$ for all $\vec\delta'\in\Delta(\Ps)$, which in part implies: $u_D(\vec{\delta^*}_1,\vec{\alpha^*}_1)\geq u_D(\vec{\delta^*}_2,\vec{\alpha^*}_1)$, and also:
 $u_A(\vec{\delta^*}_1,\vec{\alpha^*}_1)\geq u_A(\vec{\delta^*}_1,\vec\alpha')$ for all $\vec\alpha'\in\Delta(\A)$. As in the proof of Lemma~\ref{Lem:NE_is_Maximin}, referring to \eqref{eq:exptec_utilities_mixed}, the latter is equivalent to: $u_D(\vec{\delta^*}_1,\vec{\alpha^*}_1)\leq u_D(\vec{\delta^*}_1,\vec\alpha')$ for all $\vec\alpha'\in\Delta(\A)$, in particular: $u_D(\vec{\delta^*}_1,\vec{\alpha^*}_1)\leq u_D(\vec{\delta^*}_1,\vec{\alpha^*}_2)$.
 Similarly, since $(\vec{\delta^*}_2,\vec{\alpha^*}_2)$ is a NE, we have: $u_D(\vec{\delta^*}_2,\vec{\alpha^*}_2)\geq u_D(\vec\delta',\vec{\alpha^*}_2)$ for all $\vec\delta'\in\Delta(\Ps)$. Moreover:  $u_A(\vec{\delta^*}_2,\vec{\alpha^*}_2)\geq u_A(\vec{\delta^*}_2,\vec\alpha')$ for all $\vec\alpha'\in\Delta(\A)$, which equivalently means: $u_D(\vec{\delta^*}_2,\vec{\alpha^*}_2)\leq u_D(\vec{\delta^*}_2,\vec\alpha')$ for all $\vec\alpha'\in\Delta(\A)$, in particular: $u_D(\vec{\delta^*}_2,\vec{\alpha^*}_2)\leq u_D(\vec{\delta^*}_2,\vec{\alpha^*}_1)$.
 Putting the inequalities together, we get: 
 $u_D(\vec{\delta^*}_1,\vec{\alpha^*}_2)\geq u_D(\vec{\delta^*}_1,\vec{\alpha^*}_1)\geq u_D(\vec{\delta^*}_2,\vec{\alpha^*}_1)\geq u_D(\vec{\delta^*}_2,\vec{\alpha^*}_2)\geq u_D(\vec\delta',\vec{\alpha^*}_2)$ for all $\vec\delta'\in\Delta(\Ps)$. In short, $u_D(\vec{\delta^*}_1,\vec{\alpha^*}_2)\geq u_D(\vec\delta',\vec{\alpha^*}_2)$ for all $\vec\delta'\in\Delta(\Ps)$, or $\vec{\delta^*}_1$ is a picker's best response to $\vec{\alpha^*}_2$. 
 
 Similarly, we show that $\vec{\alpha^*}_2$ is a guesser's best response to $\vec{\delta^*}_1$, i.e., $u_A(\vec{\delta^*}_1,\vec{\alpha^*}_2)\geq u_A(\vec{\delta^*}_1,\vec\alpha')$ for all $\vec\alpha'\in\Delta(\A)$.
 Again, referring to \eqref{eq:exptec_utilities_mixed}, this is equivalent to showing: 
 $u_D(\vec{\delta^*}_1,\vec{\alpha^*}_2)\leq u_D(\vec{\delta^*}_1,\vec\alpha')$ for all $\vec\alpha'\in\Delta(\A)$.
Since $\vec{\delta^*}_2$ is a best response to $\vec{\alpha^*}_2$, we have: $u_D(\vec{\delta^*}_1,\vec{\alpha^*}_2)\leq u_D(\vec{\delta^*}_2,\vec{\alpha^*}_2)$.
Also, since $\vec{\alpha^*}_2$ is a best response to $\vec{\delta^*}_2$, we have: 
$u_A(\vec{\delta^*}_2,\vec{\alpha^*}_2)\geq u_A(\vec{\delta^*}_2,\vec{\alpha^*}_1)$, or 
equivalently, $u_D(\vec{\delta^*}_2,\vec{\alpha^*}_2)\leq u_D(\vec{\delta^*}_2,\vec{\alpha^*}_1)$.
We also have: $u_D(\vec{\delta^*}_2,\vec{\alpha^*}_1)\leq u_D(\vec{\delta^*}_1,\vec{\alpha^*}_1)$ because $\vec{\delta^*}_1$ is a best response to $\vec{\alpha^*}_1$.
Finally, we have $u_A(\vec{\delta^*}_1,\vec{\alpha^*}_1)\geq u_A(\vec{\delta^*}_1,\vec\alpha')$ for all $\vec\alpha'\in\Delta(\A)$, or equivalently, $u_D(\vec{\delta^*}_1,\vec{\alpha^*}_1)\leq u_D(\vec{\delta^*}_1,\vec\alpha')$ for all $\vec\alpha'\in\Delta(\A)$, since  $\vec{\alpha^*}_1$ is a best response to $\vec{\delta^*}_1$. Putting all of these inequalities together, we have:
$u_D(\vec{\delta^*}_1,\vec{\alpha^*}_2)\leq u_D(\vec{\delta^*}_2,\vec{\alpha^*}_2)\leq u_D(\vec{\delta^*}_2,\vec{\alpha^*}_1)\leq u_D(\vec{\delta^*}_1,\vec{\alpha^*}_1)\leq u_D(\vec{\delta^*}_1,\vec\alpha')$ for all $\vec\alpha'\in\Delta(\A)$. In short, $u_D(\vec{\delta^*}_1,\vec{\alpha^*}_2)\leq u_D(\vec{\delta^*}_1,\vec\alpha')$ 
or equivalently, $u_A(\vec{\delta^*}_1,\vec{\alpha^*}_2)\geq u_A(\vec{\delta^*}_1,\vec\alpha')$ for all $\vec\alpha'\in\Delta(\A)$. Therefore, putting the two observations together, $(\vec{\delta^*}_1,\vec{\alpha^*}_2)$ is a NE. 
 Establishing $(\vec{\delta^*}_2,\vec{\alpha^*}_1)$ is a NE follows similar steps.
 \end{IEEEproof}

\section{Proof of Corollary~\ref{Corr:Interchangeability_II}: Capped-Guesses NE Interchangeability II}\label{Appendix:Proof_of_Corr:Interchangeability_II}
\begin{IEEEproof}
Proof by contradiction: Suppose not, and there exist two NE 
$(\vec{\delta^*}_1,\vec{\alpha^*}_1)$ and $(\vec{\delta^*}_2,\vec{\alpha^*}_2)$ that yield the picker 
two different expected utilities $u_{D1}$ and $u_{D2}$. Without loss of 
generality, assume $u_{D1}<u_{D2}$. 
Similar to the arguments in the proof of Lemma~\ref{Lem:NE_is_Maximin}, from \eqref{eq:exptec_utilities_mixed} and the NE property for $\vec{\alpha^*}_1$, we have:
$u_D(\vec{\delta^*}_1,\vec{\alpha^*}_1)=\underline{u_D}(\vec{\delta^*}_1)$. Similarly, $u_D(\vec{\delta^*}_2,\vec{\alpha^*}_2)=\underline{u_D}(\vec{\delta^*}_2)$. 
From Lemma~\ref{Lem:NE_is_Maximin}, both $\vec{\delta^*}_1$ and $\vec{\delta^*}_2$ are Maximin strategies of the picker as well.
Hence, in particular, we must have: $\underline{u_D}(\vec{\delta^*}_1)=\max_{\vec\delta\in\Delta(\Ps)}\underline{u_D}(\vec\delta)=\underline{u_D}^{\max}$. However, this cannot be, because $\underline{u_D}(\vec{\delta^*}_1)=u_{D1}<u_{D2}=\underline{u_D}(\vec{\delta^*}_2)$.
\end{IEEEproof}

\section{Proof of Corollary~\ref{Corr:SSE_NE_equivalence}: Picker's SSE-NE Equivalence in Capped-Guesses}\label{Appendix:Proof_of_Corr:SSE_NE_equivalence}
\begin{IEEEproof}
Let $(\vec{\Hat{\delta}},\vec\alpha^{\mathrm{BR}}(\cdot))$ be a SSE in a Capped-Guesses game in which the picker has the mixed strategy commitment power.
Referring to \eqref{eq:exptec_utilities_mixed}, as in the proof of Lemma~\ref{Lem:NE_is_Maximin}, we have:
$\vec\alpha^{\mathrm{BR}}(\vec\delta)\in\arg\max_{\vec\alpha\in\Delta(\A)} u_{A}(\vec\delta,\vec\alpha)$ if and only if $\vec\alpha^{\mathrm{BR}}(\vec\delta)\in\arg\min_{\vec\alpha\in\Delta(\A)} u_{D}(\vec\delta,\vec\alpha)$. Hence, $u_{D}(\vec\delta,\vec\alpha^{\mathrm{BR}}(\vec\delta))=\underline{u_D}(\vec\delta)$.
These imply that: 
$\min_{\vec\alpha\in\Delta(\A)}u_{D}(\vec{\Hat{\delta}},\vec\alpha)=$
$u_D(\vec{\Hat{\delta}},\vec\alpha^{\mathrm{BR}}(\vec{\Hat{\delta}}))=$
$\max_{\vec\delta\in\Delta(\Ps)}u_{D}(\vec\delta,\vec\alpha^{\mathrm{BR}}(\vec\delta))=$
$\max_{\vec\delta\in\Delta(\Ps)}\underline{u_D}(\vec\delta)=$
$\underline{u_D}^{\mathrm{max}}$. 
Therefore, by definition, $\vec{\Hat{\delta}}\in\Omega^{\mathrm{D}}_{\mathrm{Maximin}}$, and hence, $\Omega^{\mathrm{D}}_{\mathrm{SSE}}\subseteq\Omega^{\mathrm{D}}_{\mathrm{Maximin}}$.
Following Lemma~\ref{Lem:NE_is_Maximin}, we thus also have $\Omega^{\mathrm{D}}_{\mathrm{SSE}}\subseteq\Omega^{\mathrm{D}}_{\mathrm{NE}}$. 
For the other direction, let $\vec{\Hat{\delta}}\in\Omega^{\mathrm{D}}_{\mathrm{Maximin}}$. Define $\vec\alpha^{\mathrm{BR}}:\Delta(\Ps)\to\Delta(\A)$ such that for a given $\vec\delta\in\Delta(\Ps)$, 
$\vec\alpha^{\mathrm{BR}}(\vec\delta)\in\arg\min_{\vec\alpha\in\Delta(\A)} u_D(\vec\delta,\vec\alpha)$. In our game, the latter implies $\vec\alpha^{\mathrm{BR}}(\vec\delta)\in\arg\max_{\vec\alpha\in\Delta(\A)} u_A(\vec\delta,\vec\alpha)$, hence satisfying one of the conditions of $\vec\alpha^{\mathrm{BR}}$. This also implies that no matter how the follower breaks ties among his best responses, the utility of the leader is the same, since the best response of the follower has to be a worst response for the leader. Finally, since $\vec{\Hat{\delta}}$ is a Maximin strategy, we have: 
$\vec{\Hat{\delta}}\in\arg\max_{\vec\delta\in\Delta(\Ps)}\underline{u_D}(\vec\delta)$, where we showed $u_D(\vec\delta,\vec\alpha^{\mathrm{BR}}(\vec\delta))=\underline{u_D}(\vec\delta)$. 
Hence, $\vec{\Hat{\delta}}\in\arg\max_{\vec\delta\in\Delta(\Ps)}u_D(\vec\delta,\vec\alpha^{\mathrm{BR}}(\vec\delta))$.
These lead to the fact that $(\vec{\Hat{\delta}},\vec\alpha^{\mathrm{BR}})$ is a SSE of the game. Therefore, $\Omega^{\mathrm{D}}_{\mathrm{Maximin}}\subseteq\Omega^{\mathrm{D}}_{\mathrm{SSE}}$.
Again, from Lemma~\ref{Lem:NE_is_Maximin}, this also shows $\Omega^{\mathrm{D}}_{\mathrm{NE}}\subseteq\Omega^{\mathrm{D}}_{\mathrm{SSE}}$, completing the proof.
\end{IEEEproof}

\section{Proof of Corollary~\ref{Corr:no_commitment_value}: No Commitment Value in Capped-Guesses}\label{Appendix:Proof_of_Corr:no_commitment_value}
\begin{IEEEproof}
{The proof is straightforward following similar arguments in the proof of Corollary~\ref{Corr:SSE_NE_equivalence}: in our game, we have that
for any $\vec\delta\in\Delta$, any $\vec\alpha^{\BR}(\vec\delta)\in\arg\max u_A(\vec\delta,\vec\alpha^{\mathrm{BR}}(\vec\delta))$ also satisfies $u_D(\vec\delta,\vec\alpha^{\mathrm{BR}}(\vec\delta))=\min_{\vec\alpha\in\Delta(\A)}u_{D}(\vec\delta,\vec\alpha)$. Hence, for any SSE strategy $(\vec{\Hat{\delta}},\vec\alpha^{\BR})$, we have: $u_D(\vec{\Hat{\delta}},\vec\alpha^{\BR}(\vec{\Hat{\delta}}))=\max_{\vec\delta(\Ps)}\underline{u_D}(\vec\delta)=\underline{u_D}^{\max}$.}
\end{IEEEproof}

\section{Proof of Corollary~\ref{Corr:uniqueness}}\label{Appendix:Proof_of_Corr:Uniqueness}
\begin{IEEEproof}
{First, note that from Lemma~\ref{Lem:NE_is_Maximin}, the picker's NE strategy as given by Proposition~\ref{Prop:NE} is also a Minimax strategy, yielding $\underline{u_{D}}^{\max}=(\sum_{i=1}^JC_i|\E_i|+\lambda K)/(\sum_{i=1}^{J}|\E_i|)$ for ``ordinary'' cases. 
We show that any deviation from $\vec{\delta^*}$, denote it by $\vec\delta'$ will not be a Maximin strategy, and hence, cannot be a NE strategy of the picker. As in the proof of Proposition~\ref{Prop:NE}, we have:
$u_D(\vec\delta',\vec{\alpha^*})= -\frac{\sum_{j=1}^JC_j|\E_j| +\lambda K}{\sum_{j=1}^J |\E_j|}\sum_{p\in\cup_{i=1}^J\E_i}\vec\delta'(p)-\sum_{i=J+1}^NC_i\sum_{p\in\E_i}\vec\delta'(p)$. Following the definition of $J$ in \eqref{eq:J_Definition}, and for non-degenerate cases, we thus have: $u_D(\vec\delta',\vec{\alpha^*})<u_D(\vec{\delta^*},\vec{\alpha^*})=\underline{u_{D}}^{\max}$. Hence, $\vec\delta'$ cannot be a Maximin strategy of the picker, and thus,  following Lemma~\ref{Lem:NE_is_Maximin}, neither is it a NE strategy of hers. This establishes the necessity of the conditions in Proposition~\ref{Prop:NE} for the NE strategy of the picker. }
\end{IEEEproof}

\section{Derivation of the Strategy Space of the Guesser in Costly-Guesses Games}\label{Appendix:Formal_Model_Derivation_Costly_Guesses}
A game of Costly-Guesses is in essence a sequential game. 
The picker makes the first move as her only turn, which is a choice of her secret from $\Ps$. The next moves are only made by the guesser. 
The set of available actions to the guesser at any step is a guess of the secret, plus the option to \emph{quit} the search.
The most natural way of modeling sequential games is the \emph{extensive form} representation, i.e., with a game tree.
A history of the game is hence a sequence composed of the secret chosen by the picker, along with the sequence of the attempts of the guesser, as long as the guesser has not quited or has not made a correct guess.
The game is finite, since, even if the guesser never quits, the secret is bound to be found in finite number of steps. 

The guesser of course does not observe the action of the picker at the first step, hence, we are dealing with an extensive  game with \emph{imperfect information} \cite[Chapters 11, 12]{osborne1994course}. Note, however, that our game is one with \emph{complete information}, since both players are assumed to know about the utilities and choices of each other.
The ambiguity of the guesser at different steps can be captured by his \emph{information sets}. The information set of a player whose turn is up is a set of past histories that he is ambiguous about having happened. We assume \emph{perfect recall}, which in our game means the guesser remembers all of his past guesses. 
Then, based on his past guesses, the guesser cannot refute histories in which the secret of the picker is anything other than the guesses so far (without making any inferences on the choice of the picker). This is because all that a failed guess reveals is that the secret is not that guess, hence the picker's secret can be anywhere in the unexplored region. 

Without commitment, a pure strategy of the picker is just her pure action: $d\in\Ps$.
A pure strategy of the guesser, is a function that assigns admissible actions to each of his information sets. 
In our game, the information sets of the guesser can be fully determined by his past actions. This is because the picker makes a move only once that is at the beginning, and that is unobserved by the guesser. Note that the sequence of past actions of the guesser cannot include ``quitting'' since the search had stopped at that point (the game had reached a terminal node). Also, we can safely ignore strategies that involve trying a guess more than once, since they are strictly dominated by removing those multiple tries of the same guess. Moreover, in any extensive form game, a strategy of a player need not determine actions for information sets that are not reachable given a player's \emph{own} earlier actions in that strategy, since such assignments are irrelevant. Specifically,  irrespective of the strategy of other players, such information sets are never reached and that plan of action is never called upon. Therefore, all strategies that assign arbitrary actions to such information sets can be lumped together and represented by one equivalent strategy, as they all represent the same meaningful plan of action. These are sometimes referred to as \emph{reduced} pure strategies. In our game, in particular, a guesser's strategy that tries $p_1$ as a first guess does not need to specify an action for the information set corresponding to a failed try of $p_2$ as the first guess.
Hence, a pure strategy of the guesser can be represented as a sequence of unique secrets from $\Ps$. 
Specifically, a pure strategy $A=\ang{a_1,\ldots,a_{\tau}}$ is equivalent to: $\big(I_{\emptyset}\to a_1, I_{\ang{a_1}}\to a_2, \ldots,I_{\ang{a_1,\ldots,a_{\tau-1}}}\to a_{\tau}, I_{\ang{a_1,\ldots,a_{\tau}}}\to \texttt{quit} \big)$, where $I_{\ang{a_1,\ldots,a_t}}$ represents the information set corresponding to the following partition of game history: $\{\ang{d,a_1,\ldots,a_t}|d\in\Ps\smallsetminus \{a_1,\ldots,a_t\}\}$.
Note the difference between a strategy profile, and an \emph{outcome} of the game: a strategy profile must provide a plan of action (a prescription) if any information set is reached, even if those information sets are never in fact reached given the strategy of the others. For instance, if the picker chooses $p_1$, and the guesser tries $p_1$ as his first attempt, the strategy of the guesser still needs to specify what he would have picked if his first try would have failed. 

\section{Proof of Lemma~\ref{Lem:uniform_the_worst}: Uniform Makes the Worst Case for the Guesser  in Costly-Guesses}\label{Appendix:Proof_of_Lem:uniform_the_worst}
\begin{IEEEproof}
We start by establishing a basic lemma. The lemma sheds light on the nature of a best response of the guesser, given a mixed strategy of the picker. Specifically, it provides three ways of improving (though not necessarily strictly) a given (pure) strategy of the guesser. 
\begin{lemma}\label{Lem:Seeking_Path_Improvement}
  In a Costly-Guesses game, consider a picker's mixed strategy $\vec\delta\in\Delta\Ps$, and a guesser's pure strategy $A=\ang{a_1,\ldots, a_T}$, where $T\geq 1$. Then each of the following procedures improves $A$ for the guesser:
 \begin{enumerate}[label=(\alph*)]
   \item {Re-ordering:} If $a_i,a_j\in A$ where $i<j$, and $\vec\delta(a_j)\geq \vec\delta(a_i)$, then $u_A(\vec\delta, A')\geq\! u_A(\vec\delta,A)$, where $A'$ is the pure strategy that is derived from swapping the positions of $a_i$ and $a_j$ in $A$ and keeping everything else the same. The inequality is strict if and only if $\vec\delta(a_j)> \vec\delta(a_i)$.
  In words, it is always better for the guesser to try a position that has received a higher probability by the picker \emph{before} a secret that has received a lower probability.
  \item {Replacement:} If $a_i\in A$ and $a_j\not\in A$, and $\vec\delta(a_j)\geq \vec\delta(a_i)$, then $u_A(\vec\delta, A')\geq u_A(\vec\delta,A)$, where $A'$ is the pure strategy that is derived from replacing $a_i$ with $a_j$ in $A$ and keeping everything else the same. The inequality is strict if and only if $\vec\delta(a_j)> \vec\delta(a_i)$.
  \item {Padding:} For a given $\vec\delta$, if $u_A(\vec\delta,A)\geq u_A(\vec\delta,A')$ where $A$ coincides with $A'$ by removing $a_t$ from it, and $\vec\delta(a_{t'})=\vec\delta(a_t)$, then $u_A(\vec\delta,A'')\geq u_A(\vec\delta,A')$, where $A''$ coincides with $A$ by removing  $\ang{a_{t'},a_{t}}$ from it. The inequality is strict if $\vec\delta(a_{t})>0$. In words, if it is beneficial for the guesser to start exploring a batch of secrets with the same probability of selection by the picker, then it is even better to explore a step further through that batch (and by repeating the argument, continue through to explore all of that batch). 
  \end{enumerate} 
\end{lemma}
\begin{IEEEproof}
 Each of the above follows directly from the expression of the expected utility of the guesser as expressed in \eqref{eq:expected_utility_pure_attacker_costly_guesses_simplified}. Specifically, for part (a), we have:
 \begin{gather*}
  u_A(\vec\delta,A')-u_A(\vec\delta,A)=  (j-i) (\vec\delta(a_j)-\vec\delta(a_i))\sigma
 \end{gather*}
For part (b) we have:
\begin{multline*}
 u_A(\vec\delta,A')-u_A(\vec\delta,A)= \\
 (\vec\delta(a_j)-\vec\delta(a_i))\gamma  + (|A|-i)\left(\vec\delta(a_j)-\vec\delta(a_i)\right)\sigma
\end{multline*}
Finally, for part (c), consider $A=\ang{a_1,...,a_i,a_{i+1},...,a_T}$,  $A'=\ang{a_1,...,a_i,a_t,a_{i+1},...,a_T}$, and 
$A''=\ang{a_1,...,a_i,a_{t'},a_{t},a_{i+1},...,a_T}$.
Simplification of $u_A(\vec\delta,A')\geq u_A(\vec\delta,A)$ leads to $\gamma \vec\delta(a_t)-\sigma 
\left[(i-T)\vec\delta(a_t)+1-\sum_{j=1}^{i}\vec\delta(a_j)\right]\geq 0$. Since $\vec\delta(a_{t'})=\vec\delta(a_{t})$, 
this can also be written as: $\gamma \vec\delta(a_{t'})-\sigma 
\left[(i-T)\vec\delta(a_{t'})+1-\sum_{j=1}^{i}\vec\delta(a_j)\right]\geq 0$. Because $\sigma\vec\delta(a_t)\geq0$, we have: $\gamma \vec\delta(a_{t'})-\sigma 
\left[(i-T)\vec\delta(a_{t'})+1-\sum_{j=1}^{i}\vec\delta(a_j)-\vec\delta(a_t)\right]\geq 0$ (strictly if $\vec\delta(a_{t})>0$). This is equivalent to $u_A(\vec\delta,A'')\geq u_A(\vec\delta,A')$.
\end{IEEEproof}

Now, going back to the proof of Lemma~\ref{Lem:uniform_the_worst}, 
first note that $\max_{\vec\alpha\in\Delta(\A)}u_A(\vec\delta,\vec\alpha)\geq 0$, since $A=\ang{\texttt{quit}}$ yields a utility of zero for the guesser. 
Following part (c) of Lemma~\ref{Lem:Seeking_Path_Improvement} and Lemma~\ref{Lem:Expected_Seeker_Over_Uniform}, we have: 
$\max_{\vec\alpha\in\Delta(\A)}u_A(\mathrm{unif}(\E),\vec\alpha)=\max\{0,\gamma -\sigma(|\E|+1)/2\}$.
Without loss of generality (through relabeling), let $\vec\delta$ be sorted in descending order, breaking ties arbitrarily.
Then, in the light of parts (a) and (b) of Lemma~\ref{Lem:Seeking_Path_Improvement}, all we are left to show is $\max_{T\in\{1,\ldots,|\E|\}}u_A(\vec\delta,\ang{a_1,\ldots,a_T})\geq \gamma -\sigma(|\E|+1)/2$. Let $\supp(\vec\delta)$ be the support of $\vec\delta$, i.e., the set of $p$ for which $\vec\delta(p)>0$, and denote $M:=|\supp(\vec\delta)|$. We show that $u_A(\vec\delta,\ang{a_1,\ldots,a_M})\geq \gamma -\sigma(|\E|+1)/2$. From the general expression of $u_A$ in~\eqref{eq:expected_utility_pure_attacker_costly_guesses_simplified}, and the fact that 
$\sum_{i=1}^M\vec\delta(a_i)=1$, we get: 
$u_A(\vec\delta,\ang{a_1,\ldots,a_T})=\gamma-\sigma\sum_{i=1}^Mi\vec\delta(a_i)$. Because $\vec\delta(a_M)\leq \vec\delta(a_i)$ for all $i=1,\ldots,M$, we have: $\sum_{i=1}^Mi\vec\delta(a_i)\leq \sum_{i=1}^Mi\vec\delta(a_M) = \vec\delta(a_M)(M+1)M/2 \leq (M+1)/2$. The latter is true because $\vec\delta(a_M)\leq 1/M$, since otherwise, $\sum_{i=1}^M\vec\delta(a_i)\geq M a_M>1$. Therefore, $u_A(\vec\delta,\ang{a_1,\ldots,a_T})\geq \gamma-\sigma(M+1)/2\leq \gamma-\sigma(|\E|+1)/2$, since, obviously, for any $\delta\in\Delta(\E)$, we have $M\leq|\E|$.
This completes the proof of Lemma~\ref{Lem:uniform_the_worst}.
\end{IEEEproof}

\section{Proof of Proposition~\ref{Prop:Costly_Guesses_NE_big_gamma}: NE in Costly-Guesses (I)}\label{Appendix:Proof_of_Prop:Costly_Guesses_NE_big_gamma}
\begin{IEEEproof}
We prove the proposition using induction on the number of partitions, $N$.
For $N=1$, i.e., when $\Ps=\E_1$, the first part of the claim is trivially true: $\vec{\delta^*}(\E_1)=1$.
The second part of the claim, i.e., $u_D(\vec{\delta^*},\vec{\alpha^*})=-C_1-\lambda$ is a consequence of the following lemma:
\begin{lemma}\label{Lem:guesser_goes_all_the_way}
Consider a $\vec\delta\in\Delta(\Ps)$,  and let $\E=\supp(\vec\delta)$. 
If $\gamma>(|\E|+1)\sigma/2$, then any best response of the guesser exhaustively explores the set, i.e., 
$\arg\max_{A\in\Psi(\E)}u_A(\vec\delta,A)\subseteq \Perm(\E)$. 
\end{lemma}

In words, if uniform randomization of the picker over a set does not dissuade the guesser from entering the game, then all best responses of the guesser to any randomization of the picker lead to exhaustively exploring the set. That is, it is never optimal for the guesser to either abstain, or to explore just a portion of the set with any non-zero probability.
\begin{IEEEproof}
We establish this lemma through an induction on $|\E|$. For $|\E|=1$, $\E=\{p\}$, the condition of the lemma reduces to $\gamma>\sigma$, and hence the claim trivially follows since 
$u_A(p,\ang{p})=\gamma-\sigma>0=u_A(p,\ang{\texttt{quit}})$. Now, suppose the lemma holds for $|\E|=1,\ldots,M-1$. 
We show that it is also true for $|\E|=M$. 
First, following Lemma~\ref{Lem:uniform_the_worst}, we have:
$\max_{\vec\alpha\in\Delta(\A)}u_{A}(\vec{\delta},\vec{\alpha})\geq \max_{\vec\alpha\in\Delta(\A)}u_A(\mathrm{unif}(\E),\vec\alpha)\geq u_A(\mathrm{unif}(\E),A\in\mathrm{Perm}(\E))=\gamma-(|\E|+1)\sigma/2>0$. 
This implies that $\max_{\vec\alpha\in\Delta(\A)}u_{A}(\vec{\delta},\vec{\alpha})>u_{A}(\vec{\delta},\ang{\texttt{quit}})$.
Hence, for all $A'\in\arg\max_{A\in\A} u_A(\vec\delta, A)$, we must have $|A'|\geq 1$.  Consider an $A'\in\arg\max_{A\in\A} u_A(\vec\delta,A)$.
Because $|A'|\geq 1$, $\exists \E'\subseteq\E$ such that $\E'\neq \emptyset$ and $A'\in\Perm(\E')$. If $\E'=\E$, the induction step is proved. 
Suppose it was not the case, and $|\E'|=m$,  where $1\leq m<M=|\E|$. 
Consider the set $\E'':=\E\smallsetminus\E'$ and the following picking distribution over it: $\vec{\Hat{\delta}}(p)={\vec\delta(p)}/\left({1-\vec\delta(\E')}\right)$ for all $p\in\E''$ and $\vec{\Hat{\delta}}(p)=0$ for all $p\not\in\E''$.
Note that $\supp(\vec{\Hat{\delta}})=\E''$, and hence $|\supp(\vec{\Hat{\delta}})|=|\E|-|\E'|=M-m<M=|\E|$.
By the condition of the lemma in the induction step, $\gamma>(|\E|+1)\sigma/2$, hence, we also have $\gamma>(|\E''|+1)\sigma/2$. Therefore, by the induction hypothesis, any (pure) best response of the guesser to $\vec{\Hat{\delta}}$ is a permutation of $\E''$. Let $A''\in\arg\max_{A\in\A} u_A(\vec{\Hat{\delta}},A)$. 
Now, consider the composite pure strategy of the guesser $\ang{A',A''}$. 
Using conditional expectation -- conditioning on the event that all of the guesses in the sequence of $A'$ fail -- or by direct manipulation of the expressions in~\eqref{eq:expected_utility_pure_attacker_costly_guesses_simplified} or \eqref{eq:expected_utility_pure_attacker_costly_guesses_simplified_alternative}, we have:
\begin{gather}\label{eq:expected_guesser_conditional_expectation}
u_A(\vec\delta,\ang{A',A''})=u_A(\vec\delta,A')+\big(1-\vec\delta(\E')\big)u_A(\vec{\Hat{\delta}},A'')
\end{gather} 
Since, in particular, $\ang{\texttt{quit}}\not\in\arg\max_{A\in\A}u_A(\vec{\Hat{\delta}},A)$, we have: $u_A(\vec{\Hat{\delta}},A'')>u_A(\vec{\Hat{\delta}},\ang{\texttt{quit}})=0$. Also, $\left(1-\vec\delta(\E')\right)>0$,  because $\supp(\vec\delta)=\E$ and $|\E'|<|\E|$. Hence, \eqref{eq:expected_guesser_conditional_expectation} implies $u_A(\vec\delta,\ang{A',A''})>u_A(\vec\delta,A')$, which contradicts $A'$ being a best response to $\vec\delta$. This completes the proof the induction step, and hence, the lemma.
\end{IEEEproof}

Resuming our proof of Proposition~\ref{Prop:Costly_Guesses_NE_big_gamma} by induction, suppose the proposition holds for any number of partitions $N=1,\ldots,M-1$. We show that it also holds for $N=M$. 
Consider a NE $(\vec{\delta^*},\vec{\alpha^*})$. If there is a partition from $\E_1$ to $\E_M$ that the picker does not assign any positive probability to any of its members, i.e., $\exists \E_i$ such that $\vec{\delta^*}(\E_i)=0$, then that partition can be ignored, and the proposition follows from the induction hypothesis. Therefore, we only need to consider the cases in which $\vec{\delta^*}(\E_i)>0$ for all $i=1,\ldots,M$. 
This implies that for each $i=1,\ldots,M$, $\exists p_i\in\E_i$ such that $\vec{\delta^*}(p_i)>0$. Since $\vec{\delta^*}$ is a best response to $\vec{\alpha^*}$,
we have: $u_D(p_i,\vec{\alpha^*})$ is the same for all $i=1,\ldots,M$. Note that $u_D(p_i,\vec{\alpha^*})=-C_i-\lambda \vec{\phi^*}(p_i)$, where 
$\vec{\phi^*}(p_i)=\sum_{A\in\A}\mathbf{1}_{A}(p_i)\vec{\alpha^*}(A)$, is the aggregate probability that $p_i$ is found if the picker chooses $p_i$ as her pure strategy and the guesser chooses the mixed response of $\vec{\alpha^*}$. Firstly, we deduce that for this case to happen $C_M$ has to be less than $C_1+\lambda$, since otherwise,
$u_D(p_1,\vec{\alpha^*})=u_D(p_M,\vec{\alpha^*})$ would imply $\vec{\phi^*}(p_M)<0$, which is impossible. Hence, by the condition of the proposition in the induction step, we must have: $\gamma>(|\E_M|+1)\sigma/2$. 

Since $C_i$'s are strictly increasing in $i$, the fact that $u_D(p_i,\vec{\alpha^*})$ is the same for all $i$ means that $\vec{\phi^*}(p_i)$'s are strictly decreasing in $i$. 
Specifically, consider $i,j$ where $1\leq i\leq M-1$ and $i<j\leq M$. From $u_D(p_i,\vec{\alpha^*})=u_D(p_j,\vec{\alpha^*})$, we get $\vec{\phi^*}(p_i)-\vec{\phi^*}(p_j)=(C_j-C_i)/\lambda>0$, i.e., $\sum_{A\in\A}[\mathbf{1}_{A}(p_i)-\mathbf{1}_{A}(p_j)]\vec{\alpha^*}(A)>0$. Therefore, $\exists A'\in\supp(\vec{\alpha^*})$ such that $p_i\in A'$ but $p_j\not\in\A'$. Following part (b) of Lemma~\ref{Lem:Seeking_Path_Improvement}, this in turn implies that $\vec{\delta^*}(p_i)\geq \vec{\delta^*}(p_j)$ (namely, because otherwise, $A'$ is strictly dominated by a strategy that is achieved by replacing $p_i$ with $p_j$ in it, and hence, cannot receive a strictly positive probability in the best response of the guesser). Since the choice of $i,j$, as long as $1\leq i\leq M-1$ and $i<j\leq M$, was arbitrary, we have: 
$\vec{\delta^*}(p_i)$ is non-increasing over $i=1,\ldots,M$. 

Let $p'_i$ be an arbitrary member of $\E_i$ other than $p_i$ (if exists), where $i\in\{1,\ldots,M-1\}$.  
Suppose $\vec{\delta^*}(p'_i)=0$. Then we must have $\vec{\phi^*}(p'_i)\leq\vec{\phi^*}(p_M)$, since there is no $A\in\supp(\vec{\alpha^*})$ for which $p'_i\in A$ but $p_M\not\in A$. This is because, again according to part (b) of Lemma~\ref{Lem:Seeking_Path_Improvement}, any such strategy can be strictly improved by replacing $p_M$ with $p'_i$ in it, and hence, is strictly dominated for the guesser. 
On the other hand, since we have: $u_D(p'_i,\vec{\alpha^*})=-C_i-\vec{\phi^*}(p'_i)$ and $u_D(p_M,\vec{\alpha^*})=-C_M-\vec{\phi^*}(p_M)$, the inequality of $\vec{\phi^*}(p'_i)\leq\vec{\phi^*}(p_M)$ along with $C_i<C_M$ lead to: $u_D(p'_i,\vec{\alpha^*})>u_D(p_M,\vec{\alpha^*})$. This, however, violates $\vec{\delta^*}(p_M)>0$, because $p'_i$ strictly dominates $p_M$ for the picker. This means $\vec{\delta^*}(p'_i)>0$. Therefore, we must in particular have: $u_D(p'_i,\vec{\delta^*})=u_D(p_i,\vec{\delta^*})$, which yields: $\vec{\phi^*}(p'_i)=\vec{\phi^*}(p_i)$. This means that for all $A\in\supp(\vec{\alpha^*})$, we have: $p_i\in A$ only if $p'_i\in A$ as well.

Recall that, from $\vec{\phi^*}(p_{M-1})>\vec{\phi^*}(p_M)$, we have: $\exists \Hat{A} \in \supp(\vec{\alpha^*})$ such that $p_{M-1}\in\Hat{A}$ and $p_{M}\not\in \Hat{A}$. 
We showed that for any $i\in\{1,\ldots,M-2\}$, $\vec{\delta^*}(p_i)\geq \vec{\delta^*}(p_{M-1})$. Now, if $\vec{\delta^*}(p_i)>\vec{\delta^*}(p_{M-1})$, then as a consequence of part (b) of Lemma~\ref{Lem:Seeking_Path_Improvement}, we must have $p_i\in \Hat{A}$ as well. Likewise, if $\vec{\delta^*}(p_i)=\vec{\delta^*}(p_{M-1})$, then due to part (c) of 
Lemma~\ref{Lem:Seeking_Path_Improvement} and the fact that $\vec{\delta^*}(p_{M-1})>0$, we must have $p_i\in\Hat{A}$ too. 
Therefore, we have $p_i\in\Hat{A}$ for all $i\in\{1,\ldots,M-1\}$. Moreover, we showed that for any $i\leq M-1$, $p_i$ is on a best response strategy of the guesser  only if any other member of $\E_i$ (if exists) is also on that strategy. Putting things together, we reach the conclusion that any $p\in\cup_{i=1}^{M-1}\E_i$ is on $\Hat{A}$ (and not $p_{M}$).

Now, consider the following mixed strategy of the picker: $\vec{\tilde\delta}(p)=\dfrac{\vec{\delta^*}(p)}{1-\vec{\delta^*}(\cup_{i=1}^{M-1}\E_i)}\mathbf{1}_{\E_M}(p)$. Note that $\supp(\vec{\tilde\delta})\subseteq\E_M$, and hence $|\supp(\vec{\tilde\delta})|\leq|\E_M|$. Given the condition of the proposition applied to the induction step, and the fact that we ruled out the possibility of $C_M>C_1+\lambda$ for this case, we have $\gamma>(|\supp(\vec{\tilde\delta})|+1)\sigma/2$.
Moreover, since $\vec{\delta^*}(p_M)>0$, $\supp(\vec{\tilde\delta})\neq\emptyset$. Hence, following Lemma~\ref{Lem:guesser_goes_all_the_way} that we just proved, any best response of the guesser to $\vec{\tilde\delta}$ exhausts $\supp(\vec{\tilde\delta})$. In particular, for any $\tilde{A}\in\arg\max_{A\in\Psi(\supp(\vec{\tilde\delta}))}u_A(\vec{\tilde\delta},A)$, we have 
$u_A(\vec{\tilde\delta},\tilde{A})>u_A(\vec{\tilde\delta},\ang{\texttt{quit}})=0$.
Now, consider the following composite strategy of the guesser: $\ang{\Hat{A},\tilde{A}}$. Following a similar argument as in the proof of Lemma~\ref{Lem:guesser_goes_all_the_way}, using either conditional expectation or a direct reorganization of \eqref{eq:expected_utility_pure_attacker_costly_guesses_simplified} or \eqref{eq:expected_utility_pure_attacker_costly_guesses_simplified_alternative}, we have:
\begin{gather*}
 u_A(\vec{\delta^*},\ang{\Hat{A},\tilde{A}})=u_A(\vec{\delta^*},\Hat{A})+\left(1-\vec{\delta^*}(\cup_{i=1}^{M-1}\E_i)\right)
 u_A(\vec{\tilde\delta},\tilde{A})
\end{gather*}
Since the second term of the above equation is strictly positive, we have: $u_A(\vec{\delta^*},\ang{\Hat{A},\tilde{A}})>u_A(\vec{\delta^*},\Hat{A})$. This, however, violates the NE requirement that $\vec{\alpha^*}$ is a best response to $\vec{\delta^*}$.
This completes the proof of the proposition.
\end{IEEEproof}

\section{Proof of Proposition~\ref{Prop:Costly_Guesses_NE_average_gamma}: NE in Costly-Guesses (II)}\label{Appendix:Proof_of_Prop:Costly_Guesses_NE_average_gamma}
\begin{IEEEproof}
As we show, this proposition can be seen as a corollary of Proposition~\ref{Prop:Costly_Guesses_NE_big_gamma}. 
If $M=1$, the proposition trivially holds. For $M>1$, suppose the claim is not true, and there exists a NE $(\vec{\delta^*},\vec{\alpha^*})$ in which $u_D(\vec{\delta^*},\vec{\alpha^*})>-C_M$. Then we must have $\vec{\delta^*}(p)=0$ for all $p\in\E_i$ where $i\geq M$. 
This is because for any such $p$, $u_D(p,\vec{\alpha^*})\leq -C_M$, which is strictly less than $u_D(\vec{\delta^*},\vec{\alpha^*})$. This means that $\supp(\vec{\delta^*})\subseteq \cup_{i=1}^{M-1}\E_i$.  
Following the condition of the proposition, we also have $\gamma>(|\E_i|+1)\sigma/2$ for all $i=1,\ldots,(M-1)$. Then following Proposition~\ref{Prop:Costly_Guesses_NE_big_gamma}, any NE leads to the picker choosing only from $\E_1$ and the guesser exhausting $\E_1$. This means that $u_D(\vec{\delta^*},\vec{\alpha^*})=-C_1-\lambda$, which implies $u_D(\vec{\delta^*},\vec{\alpha^*})<-C_M$, a contradiction. 
\end{IEEEproof}

\section{Proof of Proposition~\ref{Prop:Costly_Guesses_Stackelberg}: Picker's SSE in Costly-Guesses}\label{Appendix:Proof_of_Prop:Costly_Guesses_Stackelberg}
\begin{IEEEproof}
We consider a general mixed strategy of the picker $\vec\delta$ and investigate its properties to be a best strategy to commit to. 
First, the picker must not assign higher probabilities to costlier options. Namely, if $p\in\E_i$ and $p'\in\E_j$ where $j>i$ (so that $C_i=c(p)<c(p')=C_j$) then we must have: $\vec\delta(p)\geq \vec\delta(p')$.
To see this, suppose there exist $p$, $p'\in\Ps$ such that $c(p)<c(p')$ and $\vec\delta(p)< \vec\delta(p')$. 
According to Lemma~\ref{Lem:Seeking_Path_Improvement} part (b), since $\vec\delta(p)< \vec\delta(p')$, there is no best response of the guesser in which $p$ is explored and not $p'$, because any such strategy can be strictly improved upon for the guesser by replacing $p$ with $p'$ in it. Therefore, there are only three possibilities with respect to the outcome of the game regarding these two secrets: (1)~either neither one of them are explored by the guesser;
(2)~both of them are explored; or (3)~only $p'$ is explored. In all three of these contingencies, the picker could have saved at least $(c(p')-c(p))(\vec\delta(p')-\vec\delta(p))>0$ by switching the probabilities.

Next, we assemble necessary conditions for a best response of the guesser to any $\vec\delta$ of the picker. This helps us to restrict the space of best responses and accordingly, derive a tractable formulation for the utilities of the players.  According to Lemma~\ref{Lem:Seeking_Path_Improvement}, in any guesser's (pure) best response: (a)~secrets are not to be explored out of order of their probabilities ($\vec\delta$ values), 
(b)~a secret is not to be explored unless \emph{all} secrets with strictly higher probabilities are to be explored earlier, (c)~a secret is not to be explored unless \emph{all} secrets with the same probability are to be explored just before or just after that; and finally (d)~the order of exploration among secrets with the same probability is immaterial to the guesser.
Given $\vec\delta$, let $\Li_1(\vec\delta),\ldots,\Li_{M(\vec\delta)}(\vec\delta)$ partition $\Ps$ in decreasing order of their picker-assigned probabilities, i.e., let all the secrets with the highest $\vec\delta(p)$ constitute $\Li_1$, all the secrets with the second highest $\vec\delta(p)$ constitute $\Li_2$, and so on, to $\Li_M$. 
In the light of our discussion, any (pure) best response of the guesser to $\vec\delta$, is either $\ang{\texttt{quit}}$ or has to be from the set $\A^*(\vec\delta):=\{\ang{A_1,\ldots,A_K}|  K\in\{1,\ldots, M(\vec\delta)\},\  A_i\in\Perm(\Li_i(\vec\delta))\ \text{for all}\ i=1,\ldots,K\}$. In words, any best response of the guesser to a given picking randomization is either not attempting at all, or planning to exhaustively explore the secrets with probabilities above a corresponding threshold, before quitting. A best response of the guesser to a given $\vec\delta$ is then selected among the above set that yields him the highest expected utility. 
Note that the number of partitions $M$ as well as the partitions themselves depend on $\vec\delta$, but we may drop the explicit dependence whenever not ambiguous for brevity.

The following lemma characterizes a key property of a SSE strategy profile:
\begin{lemma}
 Let $(\vec{\delta^*},\vec\alpha^{\BR}(\cdot))$ be a SSE strategy profile of the Costly-Guesses game. Then: (A)~$\vec{\alpha}^{\BR}(\vec{\delta^*})=\ang{\texttt{quit}}$ with probability one; \textit{or} (B)~ $u_D(\vec{\delta^*},\vec\alpha^{\BR}(\vec{\delta^*}))=-C_1-\lambda$.
\end{lemma}

That is, any SSE strategy of the picker either leads to the choice of a cheapest secret and certain revelation of it, or should be a randomization that dissuades the guesser from exploring at all. In other words, if at all it is worth randomizing to protect a secret from a guesser with costly guesses, then it should be done such that the guesser is completely deterred from entering the game. 
The proof of the lemma follows:
\begin{IEEEproof}
Let the partitions $\Li_1(\vec{\delta^*}),\ldots,\Li_{M(\vec{\delta^*})}(\vec{\delta^*})$ and the corresponding guesser's best response superset $\A^*(\vec{\delta^*})$ be as we described earlier. 
Given the first property of any SSE strategy of the picker, the choosing cost of the secrets must be non-decreasing across these partitions, i.e., if $p\in\Li_{i}$ and $p'\in\Li_j$ for $i<j$, since by the definition of partitions $\vec{\delta^*}(p)>\vec{\delta^*}(p')$,  we must have: $c(p)\leq c(p')$. Hence, in particular, $\forall p\in \Li_1$, $c(p)\geq C_1$. 
We establish the (disjunctive) statement of the lemma as follows: we show that if (B) does not hold, then (A) has to hold (a.k.a. proof by disjunctive syllogism). 

Consider the cases where (B) does not hold. Since $-C_1-\lambda$ is the Maximin utility of the picker (by simply choosing from the cheapest partition), violation of (B) means:
\begin{gather}\label{eq:when_B_is_vilated}
u_D(\vec{\delta^*},\vec\alpha^{\BR}(\vec{\delta^*}))>-C_1-\lambda. 
\end{gather}
Let $A$ be a best response of the guesser to $\vec{\delta^*}$. As we argued before, either $A=\ang{\texttt{quit}}$ or $A\in\A^*(\vec{\delta^*})$, i.e., $A=\ang{A_1,\ldots,A_K}$ for a $K\in{1,\ldots M}$, where $A_i\in\Perm(\Li_i)$ for all $i=1\ldots K$. 
If in all best responses of the guesser, we have $K=M$, i.e., the guesser explores the entire $\Ps$, then $u_D(\vec{\delta^*},\vec\alpha^{\BR}(\vec{\delta^*}))\leq -C_1-\lambda$, which is a contradiction with~\eqref{eq:when_B_is_vilated}. 

Recall from~\eqref{eq:expected_guesser_conditional_expectation} that, given a mixed strategy of the picker $\vec\delta$, for any guesser's strategy $A$ that is a concatenation of two nonempty subsequences $A',A''$, i.e. $A=\ang{A',A''}$, we have the general relation that:

\small\begin{gather}\label{eq:expected_guesser_conditional_expectation_rewritten}
u_A(\vec\delta,\ang{A',A''})=u_A(\vec\delta,A')+\big(1-\sum_{p'\in A'}\vec\delta(p')\big)u_A(\vec{\Hat{\delta}},A'')
\end{gather}\normalsize
where $\vec{\Hat{\delta}}$ is the posterior (Bayesian) belief of the guesser about the secret of the picker if all the guesses in $A'$ fail, specifically: $\vec{\Hat{\delta}}(p)=\vec\delta(p)/\left(1-\sum_{p'\in\A'}\vec\delta(p')\right)$ for all $p\not\in A'$ and $\vec{\Hat{\delta}}(p)=0$ for all $p\in A$. 
Now, suppose (A) does not hold, and hence, $A=\ang{A_1,\ldots,A_K} \in\supp(\vec{\alpha}^{\BR}(\vec{\delta^*}))$  for a $K\in\{1,\ldots, M-1\}$. The fact that $\ang{A_1,\ldots,A_K}$ is a best response to $\vec{\delta^*}$ implies that it should yield the guesser at least as much utility (or better) than any alternative strategy. In particular, 
we must have $u_A(\vec{\delta^*},\ang{A_1,\ldots,A_K})\geq u_A(\vec{\delta^*},\ang{A_1,\ldots,A_{K'}})$ for all $K'\in\{K+1,\ldots,M\}$, where $A_i\in\Li_i$ for all $i=1,\ldots K'$. 
From~\eqref{eq:expected_guesser_conditional_expectation_rewritten}, we can write: 
\begin{multline*}
 u_A(\vec{\delta^*},\ang{A_1,\ldots,A_{K'}})= u_A(\vec{\delta^*},\ang{A_1,\ldots,A_K}) \\+ \big(1-\vec{\delta^*}(\cup_{i=1}^K\Li_i)\big)u_A(\vec{\Hat{\delta}},\ang{A_{K+1},\ldots,A_{K'}})
\end{multline*}
Therefore, we must have  $\big(1-\vec{\delta^*}(\cup_{i=1}^K\Li_i)\big)=0$ or $u_A(\vec{\Hat{\delta}},\ang{A_{K+1},\ldots,A_{K'}})\leq 0$ for all $K'=K+1,\ldots, M$. 
 The former means that this best response of the guesser explores the entire support of $\vec{\delta^*}$. Recall that in SSE, the follower (guesser) breaks the ties among his best responses in favor of the leader (picker). Hence, if there was any other best response of the guesser that partially explores the $\supp(\vec{\delta^*})$ (or does not explore it at all), then $\ang{A_1,\ldots,A_K}$ could not be in $\supp(\vec\alpha^{\BR}(\vec{\delta^*}))$. Hence, we have $u_D(\vec{\delta^*},\vec\alpha^{\BR}(\vec{\delta^*}))=-\sum_{p\in\Ps}\vec{\delta^*}(p)c(p) -\lambda \leq -C_1-\lambda$, which is contradicting~\eqref{eq:when_B_is_vilated}. 
 
 On the other hand, $u_A(\vec{\Hat{\delta}},\ang{A_{K+1},\ldots,A_{K'}})\leq 0$ for all $K'=K+1,\ldots, M$ implies that  $\vec\alpha^{\BR}(\vec{\Hat{\delta}})$ is just $\ang{\texttt{quit}}$  with probability one. To see this, note that  $\vec{\Hat{\delta}}$ preserves the order of probabilities across partitions $\Li_{K+1},\ldots,\Li_{M}$ and is zero over partitions $\Li_1,\ldots,\Li_K$. Hence, following Lemma~\ref{Lem:Seeking_Path_Improvement} as before, any $A\in\supp(\vec\alpha^{\BR}(\vec{\Hat{\delta}}))$ can be written as $\ang{A_{K+1},\ldots,A_{K'}}$ for a $K'\leq M$, in which $A_i\in\Perm(\Li_i)$ for all $i$.  
 The utility of the guesser to any of such strategies is non-positive. We also have $u_A(\vec{\Hat{\delta}},\ang{\texttt{quit}})=0$. Hence, $\ang{\texttt{quit}}\in\supp(\vec\alpha^{\BR}(\vec{\Hat{\delta}}))$. Moreover, since in a SSE, the follower breaks ties in favor of the leader, there is no other strategy in the support of $\vec\alpha^{\BR}(\vec{\Hat{\delta}})$, since any other strategy of the guesser involves exploring a subset of the support of $\vec{\Hat{\delta}}$, which strictly reduces the utility of the picker.
 Therefore: 
 \begin{gather}\label{eq:u_D_delta_hat_costly_guesses}
u_D(\vec{\Hat{\delta}},\vec\alpha^{\BR}(\vec{\Hat{\delta}}))=-\sum_{p\in\cup_{i=K+1}^M\Li_i}c(p)\vec{\Hat{\delta}}(p)  
 \end{gather}
 We show that $u_D(\vec{\delta^*},\vec\alpha^{\BR}(\vec{\delta^*}))<u_D(\vec{\Hat{\delta}},\vec\alpha^{\BR}(\vec{\Hat{\delta}}))$, contradicting the assumption that  $\vec{\delta^*}$ is a SSE strategy of the picker.
 First, because $\ang{A_1,\ldots,A_K}\in\supp(\vec\alpha^{\BR}(\vec{\delta^*}))$, and since the guesser breaks his ties in favor of the picker, any strategy in $\supp(\vec\alpha^{\BR}(\vec{\delta^*}))$ explores the whole partitions of $\Li_1,\ldots,\Li_K$ as well. Hence:
 \begin{gather}\label{eq:u_D_delta_star_costly_guesses}
  u_D(\vec{\delta^*},\vec\alpha^{\BR}(\vec{\delta^*})) = -\sum_{p\in\Ps} c(p) \vec{\delta^*} (p) - \lambda  \vec{\delta^*}(\cup_{i=1}^K \Li_i)
 \end{gather}
We can expand: $\sum_{p\in\Ps} c(p) \vec{\delta^*} (p)$ as $\sum_{p\in \cup_{i=1}^K \Li_i} c(p) \vec{\delta^*} (p)+\sum_{p\in\cup_{i=K+1}^M\Li_i} c(p) \vec{\delta^*} (p)$. For the first term, we have:
$\sum_{p\in \cup_{i=1}^K \Li_i} c(p) \vec{\delta^*} (p)\geq C_1\sum_{p\in \cup_{i=1}^K \Li_i} \vec{\delta^*} (p)$. Hence, \eqref{eq:u_D_delta_star_costly_guesses} leads to:
\begin{gather*}
  u_D(\vec{\delta^*},\vec\alpha^{\BR}(\vec{\delta^*})) \leq (-C_1-\lambda)\vec{\delta^*}(\cup_{i=1}^K \Li_i)\! -\!\!\!\!\!\!\!\!\!\sum_{p\in\cup_{i=K+1}^M\Li_i}\!\!\!\!\!\!\! c(p) \vec{\delta^*} (p)
\end{gather*}
Because of \eqref{eq:when_B_is_vilated}, and since $\sum_{p\in \cup_{i=1}^K \Li_i} \vec{\delta^*}(p)>0$ (otherwise $A=\ang{A_1,\ldots,A_K}$ cannot be a best response to $\vec{\delta^*}$), the above inequality further gives:
\small
\begin{gather*}
  u_D(\vec{\delta^*}\!\!,\vec\alpha^{\BR}(\vec{\delta^*})) < u_D(\vec{\delta^*}\!\!,\vec\alpha^{\BR}(\vec{\delta^*}))
  \vec{\delta^*}(\cup_{i=1}^K \Li_i) -\!\!\!\!\!\!\!\!\!\sum_{p\in\cup_{i=K+1}^M\Li_i}\!\!\!\!\!\!\!\!\! c(p) \vec{\delta^*} (p)
\end{gather*}\normalsize
which means:
\begin{gather*}
  u_D(\vec{\delta^*},\vec\alpha^{\BR}(\vec{\delta^*})) < -\!\!\!\sum_{p\in\cup_{i=K+1}^M\Li_i}\!\! c(p) \frac{\vec{\delta^*} (p)}{\sum_{p'\in \cup_{i=1}^K \Li_i} \vec{\delta^*}(p')}
\end{gather*}
The right hand side is exactly equal to $u_D(\vec{\Hat{\delta}},\vec\alpha^{\BR}(\vec{\Hat{\delta}}))$ as given in \eqref{eq:u_D_delta_hat_costly_guesses}. This completes the proof of the lemma.
%
\end{IEEEproof}

The lemma establishes that a SSE strategy of the picker is the ``cheapest'' randomization of her that removes any ``strict'' preference of the guesser to enter the game (if the guesser is indifferent whether to enter the game it is assumed that he does not, since he breaks ties in favor of the picker in any SSE). Next, we show that the picker can restrict her search for the cheapest deterring randomizations to only among those that assign the same probability within a partition of the same cost (i.e., same probability to all members of a given $\E_i$). To see this, suppose an optimal distribution $\vec{\delta^*}$ indeed violates this property. Consider an $\E_I$ over which, the distribution is assigning distinct probabilities, i.e., $\exists p,\ p' \in \E_I$ where $\vec{\delta^*}(p)\neq \vec{\delta^*}(p')$. Now consider an alternative distribution $\vec{\tilde{\delta}}$ as the following: $\vec{\tilde\delta}(p)=\vec{\delta^*}$ for all $p\not\in\E_I$ and $\vec{\tilde\delta}(p)=\vec{\delta^*}(\E_I)/|\E_I|$ for all $p\in\E_I$. We show that this alternative distribution also dissuades the guesser from entering the game, and hence, provides the same utility for the picker (because the total probability over $\E_I$ is the same). 
Consider the following partitioning of $\Ps$: $\Ps_1= \cup_{i=1}^{I-1}\E_i$, $\Ps_2=\E_I$ and $\Ps_3=\cup_{i=I+1}^N\E_i$.  
For any strategy of the guesser $A$ that only includes members $\Ps_1$, we have: $u_A(\vec{\tilde\delta},A)=u_A(\vec{\tilde\delta},A)$.
For any strategy of the guesser that includes members from $\Ps_1$ and $\Ps_2$, we can reorder $A$ as $\ang{A_1,A_2}$ where $A_1$ only includes from $\Ps_1$ and $A_2$ only from $\Ps_2$ without changing the value of $u_A(\vec{\tilde\delta},A)$ or $u_A(\vec{\tilde\delta},A)$. To see this, note that: (1)~as we argued before, any SSE strategy of the picker assigns non-increasing probabilities across $\E_i$, hence $\vec{\delta^*}$ is non-increasing across $\E_i$, and by construction, so is $\vec{\tilde\delta}$; (2)~as we discussed before, any best response of the guesser tries guesses in decreasing order of their picker-assigned probabilities, and changing the order among options with the same probability does not affect his utility. Now, from the relation in~\eqref{eq:expected_guesser_conditional_expectation_rewritten} we have:
\begin{gather*}
 u_A(\vec{\delta^*},A)=u_A(\vec{\delta^*},A_1)+\big(1-\sum_{p\in A_1}\vec{\delta^*}(p)\big)u_A(\vec{\Hat{\delta}^*},A_2)
\end{gather*}
The same can be written for $u_A(\vec{\tilde\delta},A)$:
\begin{gather*}
  u_A(\vec{\tilde\delta},A)=u_A(\vec{\tilde\delta},A_1)+\big(1-\sum_{p\in A_1}\vec{\tilde\delta}(p)\big)u_A(\vec{\Hat{\tilde\delta}},A_2)
\end{gather*}
Note that $u_A(\vec{\delta^*},A_1)=u_A(\vec{\tilde\delta},A_1)$ and $\sum_{p\in A_1}\vec{\delta^*}(p)=\sum_{p\in A_1}\vec{\tilde\delta}(p)$. Moreover, since $\vec{\Hat{\tilde\delta}}=\unif(\E_I)$, from Lemma~\ref{Lem:uniform_the_worst}, we have: $u_A(\vec{\Hat{\tilde\delta}},A_2)\leq u_A(\vec{\Hat{\delta}^*},A_2)$. Hence, overall, we have: $u_A(\vec{\tilde\delta},A)\leq u_A(\vec{\delta^*},A)$. 
Finally, for any strategy of the guesser that includes from $\Ps_1$, $\Ps_2$ and $\Ps_3$, following the same argument, we can reorder $A$ as $\ang{A_1,A_2,A_3}$ such that each $A_i$ only includes from $\Ps_i$, $i=1,2,3$. Also, by using \eqref{eq:expected_guesser_conditional_expectation_rewritten} twice, noting $\big(1-\sum_{p\in A_1}\vec{\tilde\delta}(p)\big)\big(1-\sum_{p\in A_2}\vec{\Hat{\tilde\delta}}(p)\big)=\big(1-\sum_{p\in \ang{A_1,A_2}}\vec{\tilde\delta}(p)\big)$, we have:
\begin{multline*}
 u_A(\vec{\delta^*},A)=u_A(\vec{\delta^*},A_1)+\big(1-\sum_{p\in A_1}\vec{\delta^*}(p)\big)u_A(\vec{\Hat{\delta}^*},A_2)\\
 +\big(1-\sum_{p\in \ang{A_1,A_2}}\vec{\delta^*}(p)\big)u_A(\vec{\Hat{\Hat{\delta}}^*},A_3)
\end{multline*}
and similarly for $u_A(\vec{\tilde\delta},A)$. Like before, we have: $u_A(\vec{\delta^*},A_1)=u_A(\vec{\tilde\delta},A_1)$,
$u_A(\vec{\Hat{\Hat{\delta}}^*},A_3)=u_A(\vec{\Hat{\Hat{\tilde\delta}}},A_3)$ (because $\vec{\Hat{\Hat{\delta}}^*}=\vec{\Hat{\Hat{\tilde\delta}}})$, $\sum_{p\in A_1}\vec{\delta^*}(p)=\sum_{p\in A_1}\vec{\tilde\delta}(p)$, and $\sum_{p\in \ang{A_1,A_2}}\vec{\delta^*}(p)=\sum_{p\in \ang{A_1,A_2}}\vec{\tilde\delta}(p)$. Also, from (a slight extension of) Lemma~\ref{Lem:uniform_the_worst}, $u_A(\vec{\Hat{\tilde\delta}},A_2)\leq u_A(\vec{\Hat{\delta}^*},A_2)$. 
Hence, for this case, and therefore for all three cases, we have  $u_A(\vec{\tilde\delta},A)\leq u_A(\vec{\delta^*},A)$. 
This means that $u_A(\vec{\delta^*},A)\leq 0$ implies $u_A(\vec{\tilde\delta},A)\leq 0$. 

Hence, all that is left to show is that the linear programming provided in the proposition, yields the highest utility for the picker among the distributions that: (1)~assign equal values within each $\E_i$, (2)~are decreasing across $\E_i$, (3)~remove any strict incentive for the guesser to enter the game, i.e., make $\ang{\texttt{quit}}$ a best response of the guesser. 
To show each of these, note that the proposition prescribes $\nu_i/|\E_i|$ as the probability for all of the member of $\E_i$, hence it satisfies the first property. The constraints of $\nu_i\geq 0$ for all $i$, and $\sum_{i=1}^N\nu_i=1$, ensure that the solution is indeed a legitimate mixed strategy of the picker. The constraints $\nu_i/|\E_i|\geq \nu_{i+1}/|\E_{i+1}|$ for $i=1,\ldots,N-1$ ascertain the second property listed above. Finally, the remaining inequalities ensure the third property, as we show next.

Suppose the mixed strategy of the picker is $\vec\delta(p)=\nu_i/|\E_i|$ for all $p\in\E_i$, $i=1,\ldots,N$. Consider a guesser's strategy $A=\ang{A_1,\ldots,A_K}$ for some $K\in\{1,\ldots,N\}$, where $A_i\in\Perm(\E_i)$. That is, trying all the partitions thoroughly and sequentially from the cheapest partition up to partition $K$ upon failed guesses. Then the expected utility of the guesser is the following: 
$u_A(\vec\delta,A)=\gamma\sum_{i=1}^K \nu_i - \sigma\sum_{i=1}^K\left[|\E_i|(1-\sum_{j=1}^{i-1}\nu_j)-(|\E_i|-1)\nu_i/2\right]$.
There are at least two ways to obtain this equality:\\
\textbf{Method 1 -- using conditional expectation:} 
The secret is in the first partition with probability $\nu_1$. Conditioning on the secret being from partition 1, following Lemma~\ref{Lem:Expected_Seeker_Over_Uniform}, the expected utility of the guesser is  $\gamma - \sigma(|\E_1|+1)/2$. 
With probability $(1-\nu_1)$, the secret is not from partition 1. In that case, the expected utility of the guesser is $-\sigma |\E_1|$ plus the expected utility of exploring the rest of the $K-1$ partitions given that the secret is not from partition 1. Continuing this procedure, we have:
\begin{multline*}
 u_A(\vec\delta,A)\! =\! \sum_{i=1}^K \Bigg[\Big(\gamma \frac{\nu_i}{1-\sum_{j=1}^{i-1}\nu_j}-\sigma\frac{|\E_i|+1}{2}\cdot\frac{\nu_i}{1-\sum_{j=1}^{i-1}\nu_j}\\
-\sigma|\E_i|(1-\frac{\nu_i}{1-\sum_{j=1}^{i-1}\nu_j})\Big)(1-\sum_{j=1}^{i-1}\nu_j)\Bigg]
\end{multline*}
which simplifies to:
\begin{multline*}
  \gamma\sum_{i=1}^K\nu_i- \sigma\sum_{i=1}^K\left[\nu_i\frac{|\E_i|+1}{2}+|\E_i|(1-\sum_{j=1}^{i}\nu_j)\right]\\
  = \gamma\sum_{i=1}^K\nu_i- \sigma\sum_{i=1}^K\left[|\E_i|(1-\sum_{j=1}^{i-1}\nu_j)-\frac{|\E_i|-1}{2}\nu_i\right]
\end{multline*}
\textbf{Method 2 - the direct way:}
We have:
\begin{multline*}
 u_A(\vec\delta,A)= \gamma\sum_{i=1}^K\nu_i - \sigma \sum_{i=1}^K \sum_{l=1}^{|\E_i|}\left[1-\sum_{j=1}^{i-1}\nu_j-\frac{l-1}{|\E_i|}\nu_i)\right]\\
 = \gamma\sum_{i=1}^K\nu_i - \sigma \sum_{i=1}^K \left[(1-\sum_{j=1}^{i-1}\nu_j)|\E_i|-\frac{|\E_i|(|\E_i|-1)}{2}\frac{1}{|\E_i|}\nu_i)\right]
\end{multline*}
Which simplifies to the same expression as before.
This concludes the proof of the proposition. 
\end{IEEEproof}
\fi
\end{document}